\documentclass[printer]{aa}
\usepackage{graphics,epsfig,lscape,txfonts,color}
\usepackage{natbib}
\usepackage{german}
\bibpunct{(}{)}{;}{a}{}{,}

\def\cm-2{cm$^{-2}$}

\def\ein{{\it Einstein}}
\def\chandra{{\it Chandra}}

\def\xmm{{XMM-Newton}}
\def\m31{\object{M~31}}

\newcommand{\ergcm}[1]{$\times 10^{#1}$ \hbox{erg cm$^{-2}$ s$^{-1}$}}

\newcommand{\ergs}[1]{$\times 10^{#1}$ \hbox{erg s$^{-1}$}}
\newcommand{\oergs}[1]{$10^{#1}$ erg s$^{-1}$}
\newcommand{\hcm}[1]{$\times 10^{#1}$ cm$^{-2}$}

\newcommand{\nh}{\hbox{$N_{\rm H}$}}

\newcommand{\mr}{\mathrm}
\newcommand{\lb}{\left}
\newcommand{\rb}{\right}
\begin{document}
\originalTeX

   \title{Time variability of X-ray sources in the M~31 centre field
   \thanks{Based on 
   observations obtained with XMM-Newton, an ESA science mission with 
   instruments and contributions directly funded by ESA Member States 
   and NASA.}\fnmsep
   \thanks{Tables~3 and~5 are only available in electronic form
    at the CDS via anonymous ftp to cdsarc.u-strasbg.fr (130.79.128.5)
    or via http://cdsweb.u-strasbg.fr/cgi-bin/qcat?J/A+A/ }
}

   \author{H.~Stiele \and
           W.~Pietsch \and 
           F.~Haberl \and
	   M.~Freyberg 
        }
\institute{Max-Planck-Institut f\"ur extraterrestrische Physik, Giessenbachstrasse,
           85748 Garching, Germany 
           }
     
     \offprints{H.~Stiele, \email{hstiele@mpe.mpg.de}}

   \date{Received  / Accepted }
   \titlerunning{Time variability of X-ray sources in the M~31 centre field}

   	\abstract{}
        {We present an extension to our \xmm\ X-ray source catalogue of \m31\, containing 39 newly found sources. In order to classify and identify more of the sources we search for X-ray time variability in \xmm\ archival data of the \m31\ centre field.}
        {As a source list we used our extended catalogue based on observations covering the time span from June 2000 to July 2004. We then determined the flux or at least an upper limit at the source positions for each observation. Deriving the flux ratios for the different observations and searching for the maximum flux difference we determined variability factors. We also calculated the significance of the flux ratios.}
        {Using hardness ratios, X-ray variability and cross correlations with catalogues in the X-ray, optical, infrared and radio regimes, we detected three super soft source candidates, one supernova remnant and six supernova remnant candidates, one globular cluster candidate, three X-ray binaries and four X-ray binary candidates. Additionally we identified one foreground star candidate and classified fifteen sources with hard spectra, which may either be X-ray binaries or Crab-like supernova remnants in \m31\ or background active galactic nuclei. The remaining five sources stay unidentified or without classification. Based on the time variability results we suggest six sources, which were formerly classified as ``hard", to be X-ray binary candidates. The classification of one other source (XMMM31~J004236.7+411349) as a supernova remnant, has to be rejected due to the distinct time variability we found. We now classify this source as a foreground star.
	}
        {} 
	
 
\maketitle

\section{Introduction}
An ideal target for a search for time variability of X-ray sources is the bright Local Group spiral galaxy \m31\ \citep[distance 780 kpc,][]{1998AJ....115.1916H,1998ApJ...503L.131S} with its moderate Galactic foreground absorption  \citep[\nh = 7\hcm{20}, ][]{1992ApJS...79...77S}.\@  

The \ein\ X-ray observatory found 16 sources in \m31, which showed variability comparing the individual observations with each other \citep[][ hereafter TF91]{1979ApJ...234L..45V,
1990ApJ...356..119C,1991ApJ...382...82T}.\@ \citet[][ hereafter PFJ93]{1993ApJ...410..615P} compared ROSAT HRI to previous \ein\ observations and found several variable sources.
The two ROSAT PSPC surveys of \m31, covered the entire galaxy and were separated by about one year. \citet{1997A&A...317..328S,2001A&A...373...63S} found, that the intensity of 34 sources varied significantly between the observations. 

\citet{2000ApJ...537L..23G} reported on first observations of the nuclear region of \m31\ with \chandra. They found that the nuclear source has an unusual X-ray spectrum compared to the other point sources in \m31. Source catalogues, based on \chandra\ observations, of the central field of \m31\ are provided by \citet{2002ApJ...577..738K} and \citet{2002ApJ...578..114K}. Three different \m31\ disk fields, spanning a range of stellar populations, were observed by \chandra\ to compare their point source luminosity functions to that of the galaxy's bulge \citep{2003ApJ...585..298K}.\@ In a synoptic study of the disk ($\approx 0.9$ square degree) of \m31, \citet{2004ApJ...609..735W} measured the mean flux and long-term light curves for 166 objects. At least 25\% of the sources show significant variability. Bright X-ray binaries (XRBs) in globular clusters and supersoft sources (SSSs) and quasisoft sources (QSSs) were investigated by \citet[][]{2002ApJ...570..618D,2004ApJ...610..247D} and \citet{2004ApJ...610..261G}.\@ The discovery of an X-ray nova was reported by \citet{2005ApJ...620..723W}.\@ \citet{2007A&A...468...49V} used \chandra\ data to examine the low mass X-ray binaries (LMXBs) in the bulge of \m31. Good candidates for LMXBs are the so-called transient sources. Studies of transient sources in \m31\ are presented in numerous papers, e.\,g.~\citet{2006ApJ...643..356W}, \citet[][ hereafter TPC06]{2006ApJ...645..277T}, \citet{2005ApJ...632.1086W}, \citet[][ hereafter WGM06]{2006ApJ...637..479W}. Using \xmm\ and \chandra\ data, \citet{2004ApJ...616..821T} detected 43 X-ray sources coincident with globular cluster candidates from various optical surveys. They studied their spectral properties, time variability and luminosity functions. \citet{2001A&A...378..800O} used \xmm\ Performance Verification observations to study the variability of X-ray sources in the central region of \m31. They found 116 sources brighter than a limiting luminosity of 6\ergs{35} and examined the $\sim60$ brightest sources for periodic and non-periodic variability. At least 15\% of these sources appear to be variable on a time scale of several months. \citet{2003A&A...411..553B} used \xmm\ to study the X-ray binary RX J0042.6+4115 and suggested it as a Z-source. \citet{2006ApJ...643..844O} studied the population of SSSs and QSSs with \xmm. \citet{2006astro.ph.10809T} provide a study of the bright sources in the central region of \m31, including spectral properties, variability and source classification. It is based on the same \xmm\ observations analysed in this paper. Recently \citet{2007arXiv0708.0874T} reported the discovery of 217s pulsations in a bright persistent SSS.

\citet[][ hereafter PFH2005]{2005A&A...434..483P} prepared a catalogue of \m31\
point-like X-ray sources analysing all observations available at that time in the \xmm\ archive which overlap at least in part with the optical $\mr{D}_{25}$ extent of the galaxy. 
In total, they detected 856 sources. The centre part of the galaxy was covered four times with a separation
of the observations of about half a year starting in June 2000 (some of the 
regions at the boundary of the centre area are even covered five times). PFH2005 give only source properties derived from an analysis of the combined centre observations. In follow-up work (i) \citet[][]{2005A&A...430L..45P} searched for X-ray burst sources in globular cluster (GlC) sources and candidates and identified two X-ray bursters and a few more candidates, and (ii) \citet[][ hereafter PFF2005]{2005A&A...442..879P} searched for correlations with optical novae. They identified seven SSSs and one symbiotic star from the list of PFH2005 with optical novae and identified one additional \xmm\ source with an optical nova. This work was continued in \citet[][ hereafter PHS2007]{2007A&A...465..375P}. 

Similar to the \object{M~33} work of \citet{2004A&A...426...11P} PFH2005 used the hardness ratios, i.\,e.\ X-ray colours, and correlations with sources in other wavelength regimes to identify and classify the detected sources. \citet{2006A&A...448.1247M} showed, for a source population study of \object{M~33}, that X-ray flux variability on different time scales allows us to further distinguish between different source classes. Phenomena such as bursts of X-ray binaries, flares of stars or the periodic variability of pulsars occur on short time scales and can therefore be observed during one single observation. On the other hand there is long term variability. Those time scales can be covered, comparing different observations of the same source. In the field of view of \m31\ there are mainly two source classes, which are known to show strong variability (variability factor $> 10$) on time scales of years. These are X-ray binaries and SSS.\@ Among the active galactic nuclei (AGNs) narrow-line Seyfert 1 galaxies and BL Lac objects show the strongest variabilities. However only a few narrow-line Seyfert 1 galaxies are known in the entire sky, which show flux variability factors larger than 10 on time scales of half a year up to several years.  
Hence it is very unlikely that one of the strongly variable sources in \m31\ would be an AGN.

In this paper we report a search for new X-ray sources in the  
\xmm\ observations to the \m31\ centre to extend the source catalogue of PFH2005 and a time variability analysis of all the \m31\ centre sources. In Sect.~2 information about the used observations and accomplished analysis is provided. 
Sect.~3 describes the source catalogue extension. The results of the temporal variability analysis are discussed in Sect.~4. 
Discussion of the individual source classes, including X-ray identifications, are provided in Sect.~5. We draw our conclusions in Sect.~6.

\section{Observations and analysis}
For our analysis we used the archival \xmm\ observations of the central region of \m31, obtained from June 2000 to January 2002 (from observations s1 and n1 only sources which lie in the intersection with at least one of the other observations are included).\@ In addition, we analysed the July 2004 monitoring observations of the low mass X-ray binary RX J0042.6+4115 (PI Barnard), which are pointed $1.1'$ to the west of the \m31\ nucleus position. Thus we have a time span of about four years for examining time variability. Details of the observations can be found in Table~\ref{tab_obs} which shows the \m31\ field name (Col.~1), the identification number (2) and date (3) of the observation and the pointing direction (4, 5). Column 6 contains the systematic position offset. For each EPIC camera the used filter and the exposure time after screening for high background is given. To achieve comparable images and results we adapted the same background screening as in PFH2005 for the newly added observations. We had to reject ObsID 0202230301, because it shows high background throughout the observation.\@ To increase the detection sensitivity we merged the data of ObsID 0202230201, 0202230401 and 0202230501 after correction of the position offsets. The combination of these three observations is called ``b".\\

\begin{table*}
\scriptsize
\begin{center}
\caption[]{\xmm\ log of archival \m31\ observation overlapping with the optical 
$D_{25}$ ellipse (proposal numbers 010927, 011257 and 015158).}
\begin{tabular}{lllrrrlrlrlr}
\hline\noalign{\smallskip}
\hline\noalign{\smallskip}
\multicolumn{1}{c}{M 31 field} & \multicolumn{1}{c}{Obs. id.} &\multicolumn{1}{c}{Obs. dates} &
\multicolumn{2}{c}{Pointing direction} & \multicolumn{1}{c}{Offset~$^*$} & \multicolumn{2}{c}{EPIC PN} & 
\multicolumn{2}{c}{EPIC MOS1} & \multicolumn{2}{c}{EPIC MOS2}  \\ 
\noalign{\smallskip}
& & & \multicolumn{2}{c}{RA/Dec (J2000)} & \multicolumn{1}{c}{} 
& \multicolumn{1}{c}{Filter$^{+}$}  & \multicolumn{1}{c}{$T_{exp}^{\dagger}$}
& \multicolumn{1}{c}{Filter$^{+}$}  & \multicolumn{1}{c}{$T_{exp}^{\dagger}$}
& \multicolumn{1}{c}{Filter$^{+}$}  & \multicolumn{1}{c}{$T_{exp}^{\dagger}$}\\
\noalign{\smallskip}
\multicolumn{1}{c}{(1)} & \multicolumn{1}{c}{(2)} & \multicolumn{1}{c}{(3)} & 
\multicolumn{1}{c}{(4)} & \multicolumn{1}{c}{(5)} & \multicolumn{1}{c}{(6)} & 
\multicolumn{1}{c}{(7)} & \multicolumn{1}{c}{(8)} & \multicolumn{1}{c}{(9)} & 
\multicolumn{1}{c}{(10)} & \multicolumn{1}{c}{(11)} & \multicolumn{1}{c}{(12)}\\
\noalign{\smallskip}\hline\noalign{\smallskip}
Centre 1~(c1) & 0112570401 & 2000-06-25    & 0:42:36.2 & 41:16:58 & $-1.9,+0.1$ & medium  & 26.40 & medium  & 29.92 &medium  & 29.91 \\
Centre 2~(c2) & 0112570601 & 2000-12-28    & 0:42:49.8 & 41:14:37 & $-2.1,+0.2$ & medium  &  9.81 & medium  & 12.24 &medium  & 12.24 \\
Centre 3~(c3) & 0112570701 & 2001-06-29    & 0:42:36.3 & 41:16:54 & $-3.2,-1.7$ & medium  & 27.65 & medium  & 27.65 &medium  & 27.65 \\
North 1~~\,(n1) & 0109270701 & 2002-01-05    & 0:44:08.2 & 41:34:56 & $-0.3,+0.7$ & medium  & 54.78 & medium  & 57.31 &medium  & 57.30 \\
Centre 4~(c4) & 0112570101 & 2002-01-06/07 & 0:42:50.4 & 41:14:46 & $-1.0,-0.8$ & thin	& 60.79 & thin	 & 63.31 &thin	 & 63.32 \\
South 1~~\,(s1) & 0112570201 & 2002-01-12/13 & 0:41:32.7 & 40:54:38 & $-2.1,-1.7$ & thin	& 53.45 & thin	 & 53.76 &thin	 & 53.73 \\
(b1)$^{\ddagger}$ & 0202230201 & 2004-07-16 & 0:42:38.6 & 41:16:04 & $-1.3,-1.2$ & medium	& 18.30 & medium & 19.40 &medium & 19.40 \\
(b2) & 0202230301 & 2004-07-17 & 0:42:38.6 & 41:16:04 & $-1.0,-0.9$ & medium	& 0.0 & medium & 0.0 &medium & 0.0 \\           
(b3)$^{\ddagger}$ & 0202230401 & 2004-07-18 & 0:42:38.6 & 41:16:04 & $-1.7,-1.5$ & medium  & 13.80 & medium  & 17.90 &medium  & 17.90 \\
(b4)$^{\ddagger}$ & 0202230501 & 2004-07-19 & 0:42:38.6 & 41:16:04 & $-1.4,-1.8$ & medium  & 8.90 & medium  & 10.20 &medium  & 10.20 \\
\noalign{\smallskip}
\hline
\noalign{\smallskip}
\end{tabular}
\label{tab_obs}
\end{center}
Notes:\\
$^{ *~}$: Systematic offset in RA and Dec in arcsec determined from correlations with 2MASS, USNO-B1 and \chandra\ catalogues \\
$^{ +~}$: All observations in full frame imaging mode\\
$^{ {\dagger}~}$: Exposure time in units of ks after screening for high
background used for detection\\
$^{ {\ddagger}~}$: Combination of the three observations is called b (see text)
\normalsize
\end{table*}
We searched for sources in ``b", which were not visible in the X-ray wavelength regime about 2.5 years earlier. In addition we reexamine observations c1, c2, c3 and c4 individually, to search for sources not included in the PFH2005 catalogue, which -- besides source 856 -- was based on an analysis of the merged images of observations c1 to c4. \\ 

The data analysis was performed using tools in the \xmm\ Science Analysis System
(SAS) v6.6.0 and v7.0.0, 
EXSAS/MIDAS 03OCT\_EXP, and 
FTOOLS v6.0.6 software packages, the imaging application DS9 v4.0b7 together with
the funtools package,
the mission count rate simulator WebPIMMS v3.6a
and the spectral 
analysis software XSPEC v11.3.1.

\subsection{Images}
We used five energy bands: (0.2--0.5) keV, (0.5--1.0) keV, (1.0--2.0) keV, (2.0--4.5) keV, and (4.5--12) keV, to create   
images, background images and exposure maps for PN, MOS1 and MOS2 and masked them for acceptable detector area. For PN, the background
maps contain the contribution from the ``out of time (OOT)" events (parameter
{\tt withootset=true} in task {\tt esplinemap}).

Figure~\ref{ima_centre} shows logarithmically scaled \xmm\ EPIC low background images  of the \m31\ centre observations integrated in 1\arcsec$\times$1\arcsec\ pixels combining data from the PN, MOS1 and MOS2 cameras in the (0.2--4.5) keV XID band. The data are smoothed with a 2D-Gaussian of $FWHM$ 5\arcsec, which corresponds to the point spread function in the centre of the field of view (FOV). Figure~\ref{ima_zoom} gives a zoom-in to the crowded centre region. 

\subsection{Source detection}
We searched for sources using simultaneously $5\times 3$
images (5 energy bands for each EPIC camera). A preliminary source
list created with the task {\tt eboxdetect} with a low likelihood threshold ({\tt likemin = 5}) was used as starting point for the task {\tt emldetect} (v. 4.44.19).\@ We used parameters {\tt nmaxfit = 2} and {\tt fitextent = true}. 
The parameter {\tt extentmodel} was set to {\tt beta} and we only allowed multi-PSF fitting if the likelihood was larger than 10.\@ Setting the parameter {\tt withtwostage} to {\tt true} the program checked in a second run whether splitting extended sources into two point-like sources ({\tt nmulsou = 2}) would achieve a more reliable result.
   
For most sources, band 5 just adds noise to the total count rate. If converted
to  fluxes this noise often dominates the total flux due to the high ECF. To
avoid this  problem we calculated count rates and fluxes for detected sources
in the ``XID" (0.2--4.5) keV band (bands 1 to 4 combined).  While for most sources this is
a good solution, for extremely hard or  soft sources there may still be bands
just adding noise. This then may lead to rate  and flux errors that seem to
wrongly indicate a lower source significance. A  similar effect occurs for the
all instrument rates and fluxes if a source is  mainly detected in one
instrument (e.g. soft sources in PN).\\

We accepted sources which have a likelihood above 6 in the combined fit. We rejected spurious detections in the vicinity of bright sources. In regions with a highly structured background the SAS detection task {\tt emldetect} registered some extended sources. We also rejected these ``sources" as spurious detections. Two sources were added manually: Source 871 was first detected as nova WeCAPP-N2001-12 and in the POINT-AGAPE variable star catalogue\  \citep{2004MNRAS.351.1071A}.\@ \citet{2004MNRAS.351.1071A} propose the hard X-ray transient [OBT2001]~3\ \citep{2001A&A...378..800O} as a counterpart, which is source 287 in the PFH2005 catalogue. PFF2005 showed that several points speak against this identification and that a faint SSS close to the position of [PFH2005]~287, which is only visible during observation c4, is a more reliable counterpart. Source 885 (see Table~3) is clearly visible in observation b (see Fig.~\ref{ima_zoom}) and we could not find any reason, why {\tt emldetect} did not automatically find it. As the source was already reported with \chandra\ \citep[][ r2-41]{2002ApJ...577..738K}, we took the \chandra\ position to derive the source parameters, using {\tt emldetect} with fixed position.

Our source catalogue extension only contains sources not already found by PFH2005. These sources were ordered according to increasing right ascension for each observation individually. Finally we merged the source lists and numbered the sources consecutively. If a source was detected in more than one observation, we took the source parameters from the first observation, in which it was detected. As this catalogue is an extension of the source list of PFH2005, new sources start with number 857.
 
To classify the source spectra we computed four hardness ratios from the source
count rates. These hardness ratios and errors are defined as  
\begin{equation}
HRi = \frac{B_{i+1} - B_{i}}{B_{i+1} + B_{i}}\; \mbox{and}\;\; EHRi = 2  \frac{\sqrt{(B_{i+1} EB_{i})^2 + (B_{i} EB_{i+1})^2}}{(B_{i+1} + B_{i})^2},
\end{equation}
for {\it i} = 1 to 4, 
where $B_{i}$ and $EB_{i}$ denotes count rates and corresponding errors in
band {\it i} as defined above. The identification and classification criteria are given in Table~\ref{class}. The source catalogue extension is presented in Sect.~\ref{srccat} (see Table~3).
 
\subsection{Variability calculation}
To examine the time variability of each source, we determined the XID flux -- or at least an upper limit for the XID flux -- at the source position in each observation . We used the task {\tt emldetect} (v. 4.60) setting the parameter {\tt xidfixed = yes} which forced {\tt emldetect} not to alter the source positions in calculating the total flux. To obtain fluxes and upper limits for all sources in our input list we set the detection likelihood threshold to 0.
    
Merging the source catalogue extension (see Sect.~3) with the source catalogue of PFH2005 we generated a starting list for our variability analyses. This starting list only contains the number and position of each source. To give correct results the task {\tt emldetect} has to process the sources from the brightest to the faintest one. Therefore we first had to order the sources in each observation by detection likelihood. For sources not visible in the observation we arbitrarily set the detection likelihood to 1.
This list was used as input for a first {\tt emldetect} run. This way we achieved an output list, in which a detection likelihood was allocated to every source. To finally examine the sources ordered by detection likelihood, a second {\tt emldetect} run was necessary. 

We only accepted XID fluxes, which are at least three times larger than their 1 $\sigma$ errors. Otherwise the triplicated error was used as an upper limit. The largest XID flux of each source was derived, excluding upper limit values (column fmax in Table~5).\@ Comparing the XID fluxes of the different observations with each other, we calculated the significance of the difference 
\begin{equation}
S_{\mr{var}}=\lb( F_{\mr{max}}- F_{\mr{min}}\rb) /\sqrt{\sigma_{\mr{max}}^2+\sigma_{\mr{min}}^2}
\end{equation}
(column svar in Table~5) and the ratio of the XID fluxes $F_{\mr{var}}=F_{\mr{max}}/F_{\mr{min}}$ (column fvar in Table~5), if $F_{\mr{max}}$ was not an upper limit. $F_{\mr{max}}$ and $F_{\mr{min}}$ are the source XID maximum and minimum (or upper limit) flux and $\sigma_{\mr{max}}$ and $\sigma_{\mr{min}}$ are the errors of the maximum and minimum flux, respectively. The results of the time variability analyses are discussed in Sect.~4.    

\begin{figure*}
\resizebox{\hsize}{!}{\includegraphics[clip]{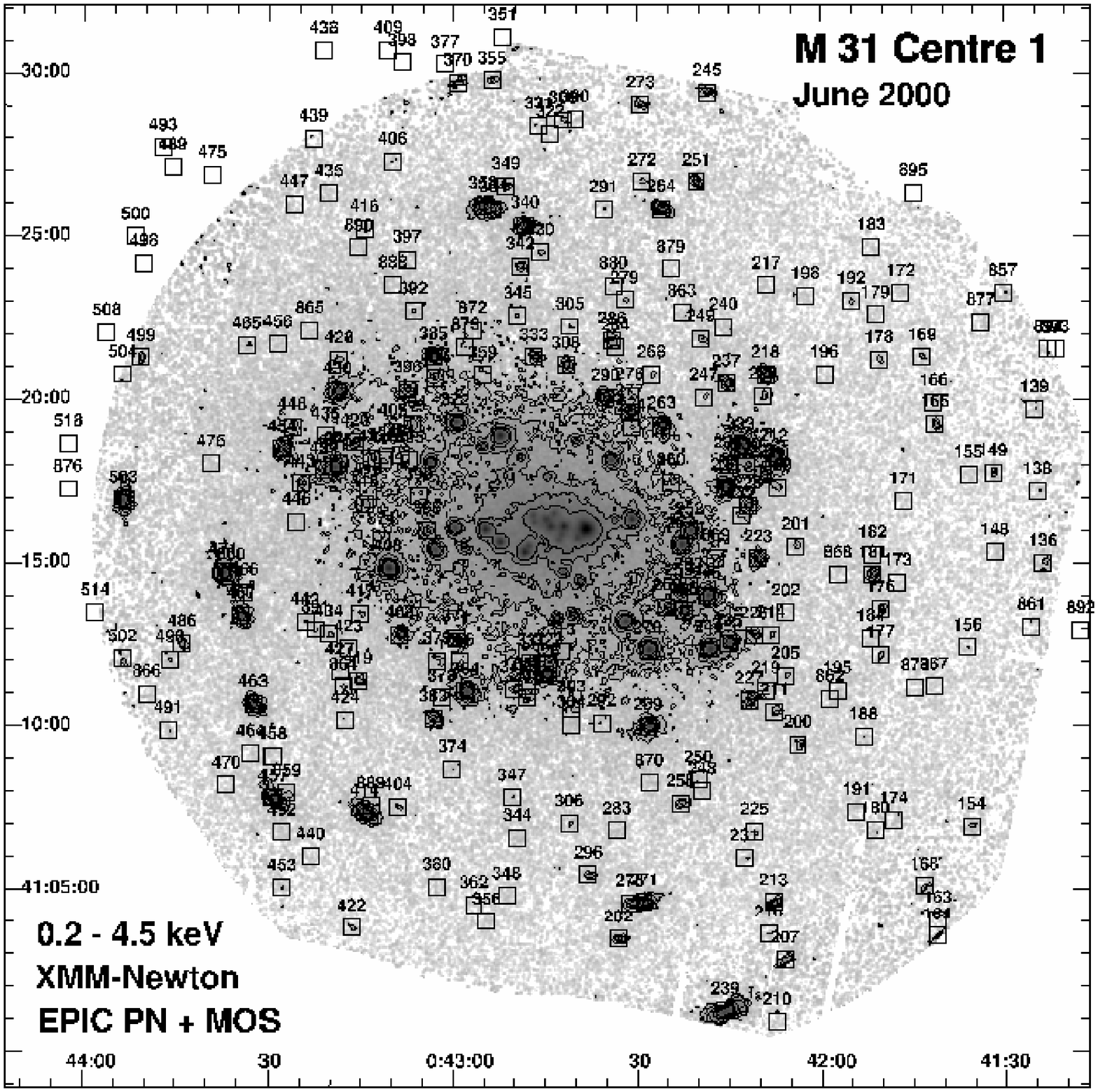}\hskip0.2cm\includegraphics[clip]{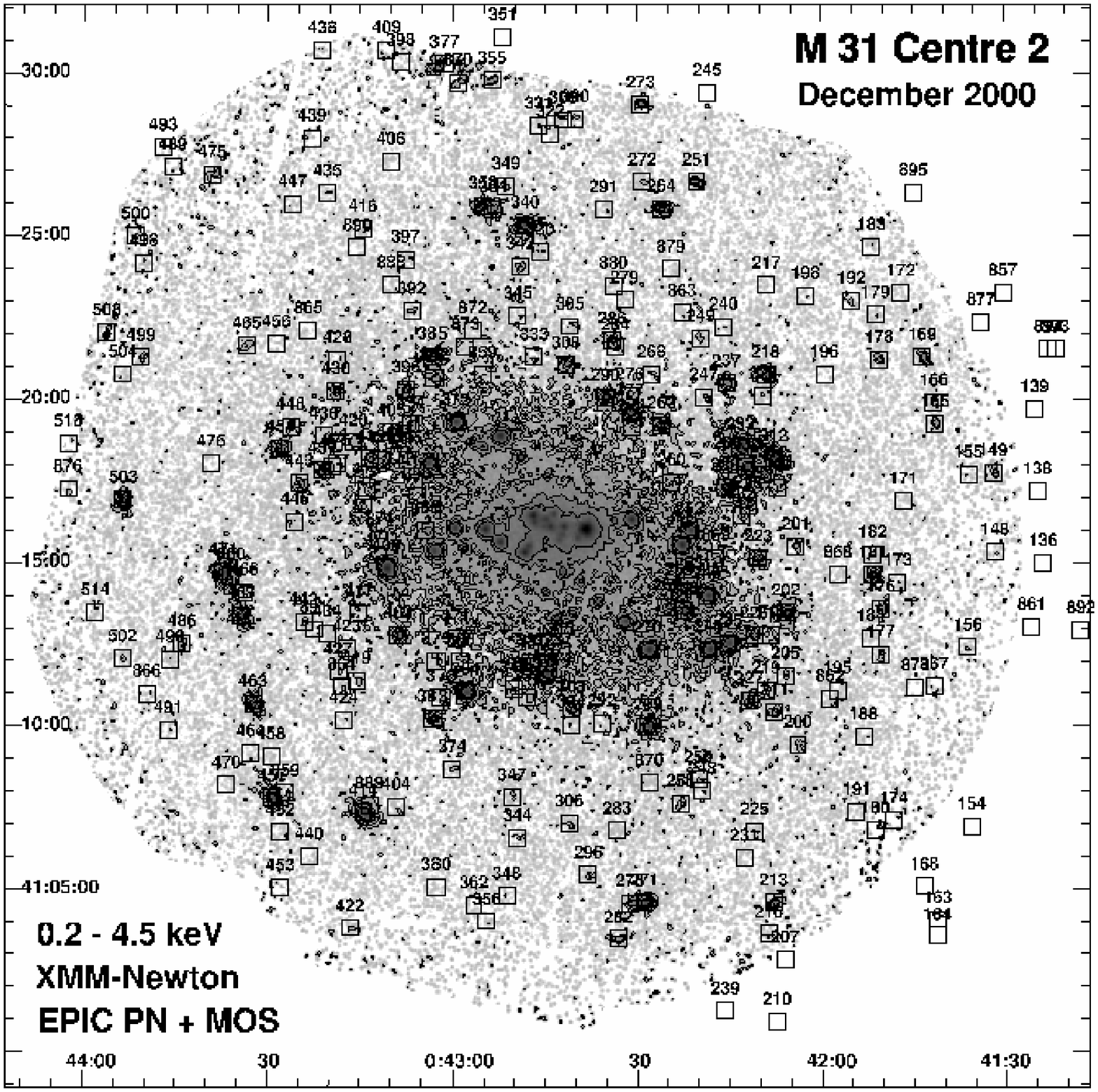}}
\resizebox{\hsize}{!}{\includegraphics[clip]{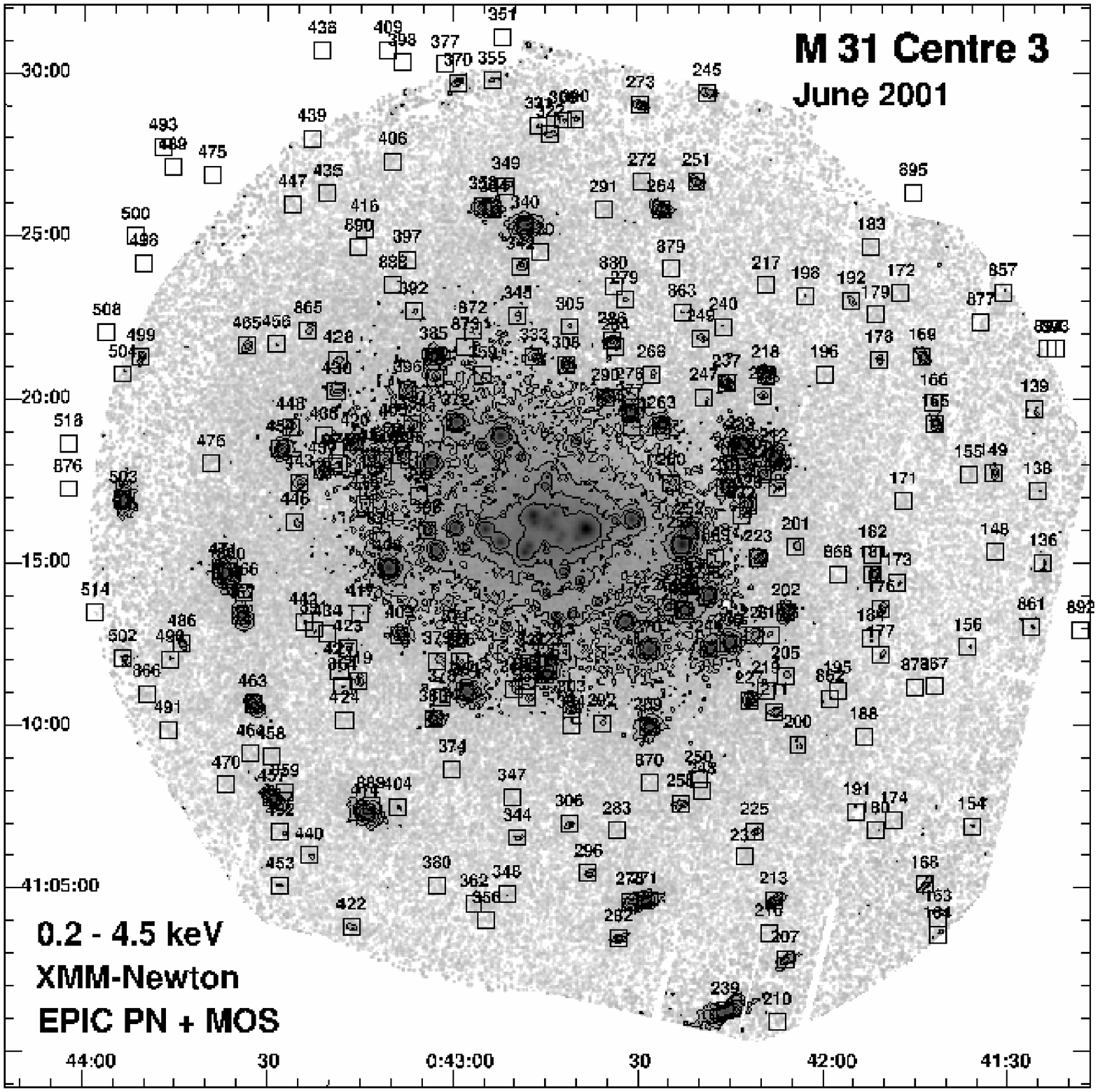}\hskip0.2cm\includegraphics[clip]{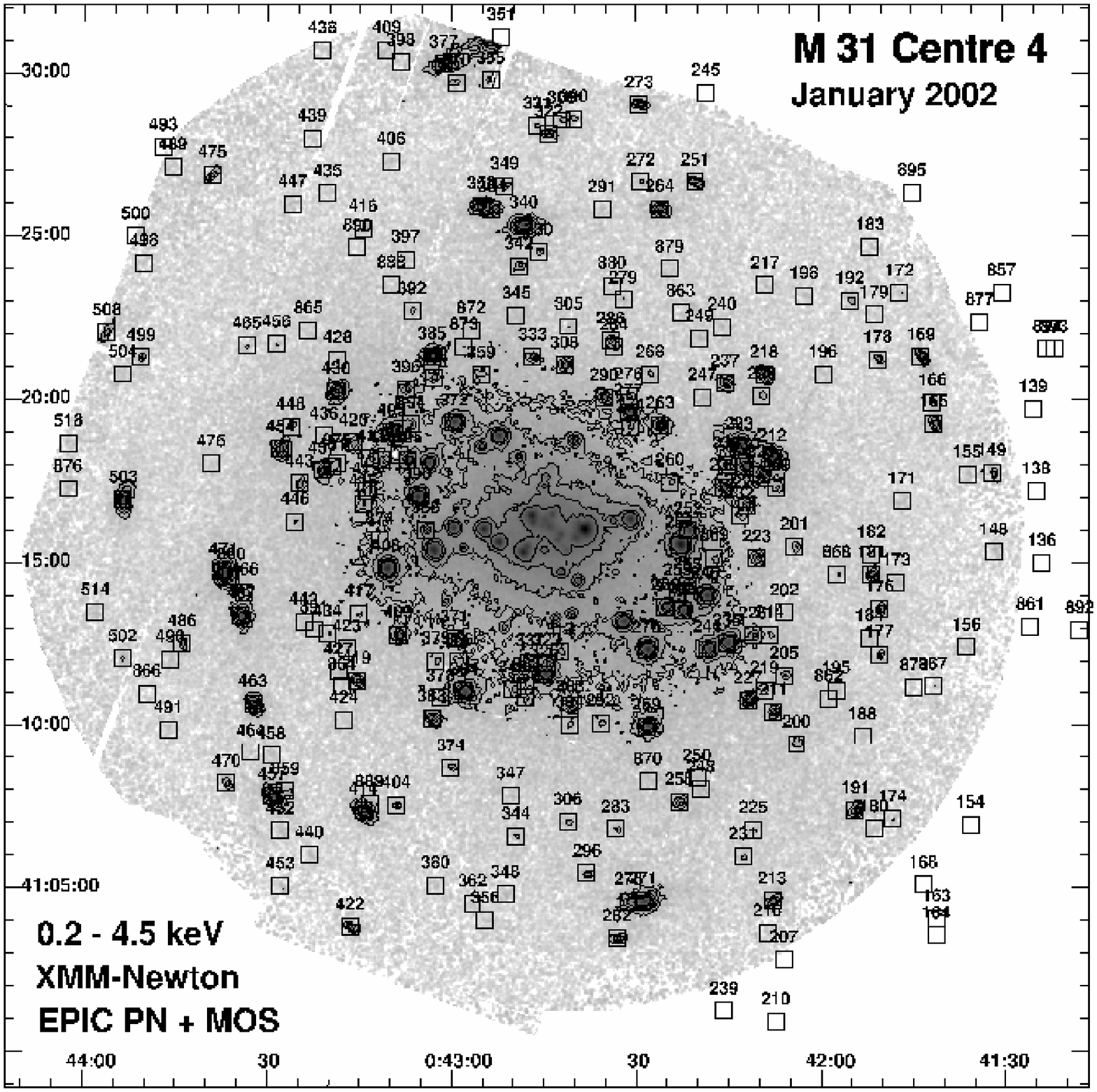}}
     \caption[]{Logarithmically scaled \xmm\ EPIC low background images
     integrated in 1\arcsec\ pixels of the \m31\ centre observations combining PN
     and MOS1 and MOS2 cameras in the (0.2--4.5) keV XID band.  
     The data are
     smoothed with a 2D-Gaussian of $FWHM$ 5\arcsec, which corresponds to the point spread function in the centre area. The images are corrected for unvignetted exposures. Contours are at $(2, 4, 8, 16, 32)\times 10^{-6}$ ct s$^{-1}$ pix$^{-1}$ including a factor of two smoothing. Sources from the combined catalogue 
      are marked in the outer area. The inner area is shown in detail in Fig.~\ref{ima_zoom}.
}
    \label{ima_centre} 
\end{figure*}

\begin{figure*}
   \resizebox{\hsize}{!}{\includegraphics[clip]{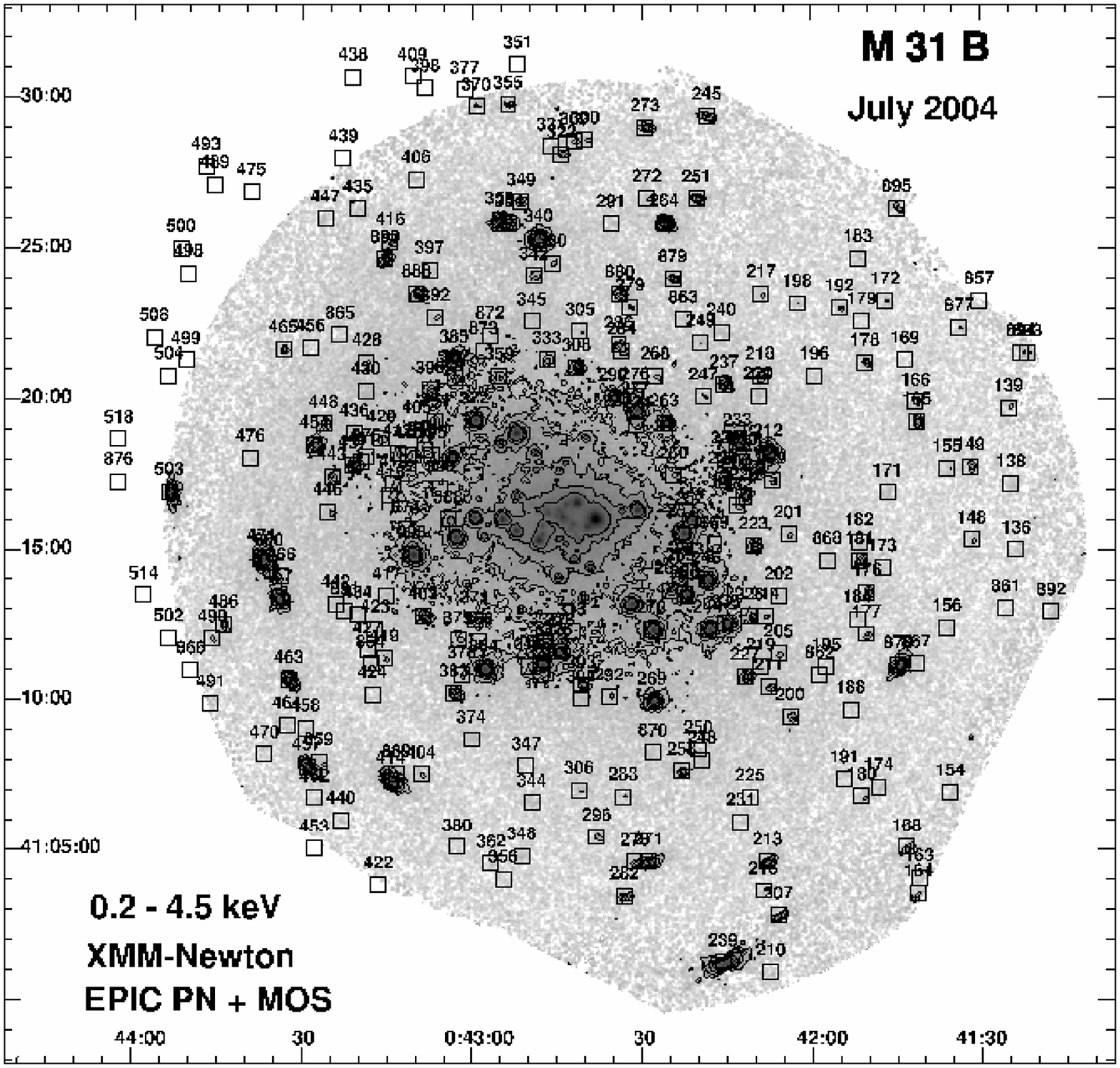}\hskip19.2cm}
\addtocounter{figure}{-1}
     \caption[]{(continued) Logarithmically scaled \xmm\ EPIC low background images
     integrated in 1\arcsec pixels of the \m31\ centre observations combining PN
     and MOS1 and MOS2 cameras in the (0.2--4.5) keV XID band. 
}
\end{figure*}

\begin{figure*}
  
\resizebox{\hsize}{!}{\includegraphics[clip]{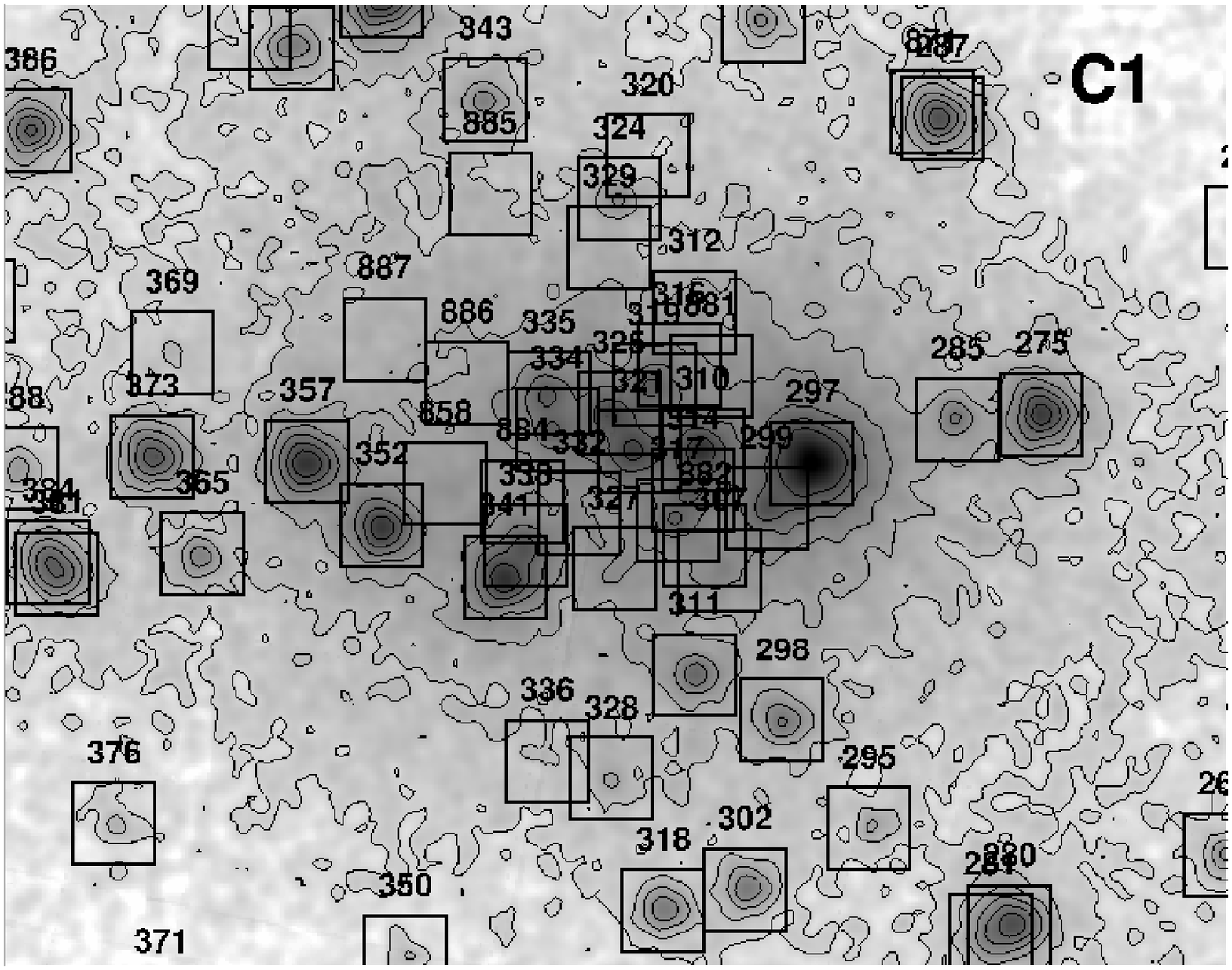}\hskip0.2cm\includegraphics[clip]{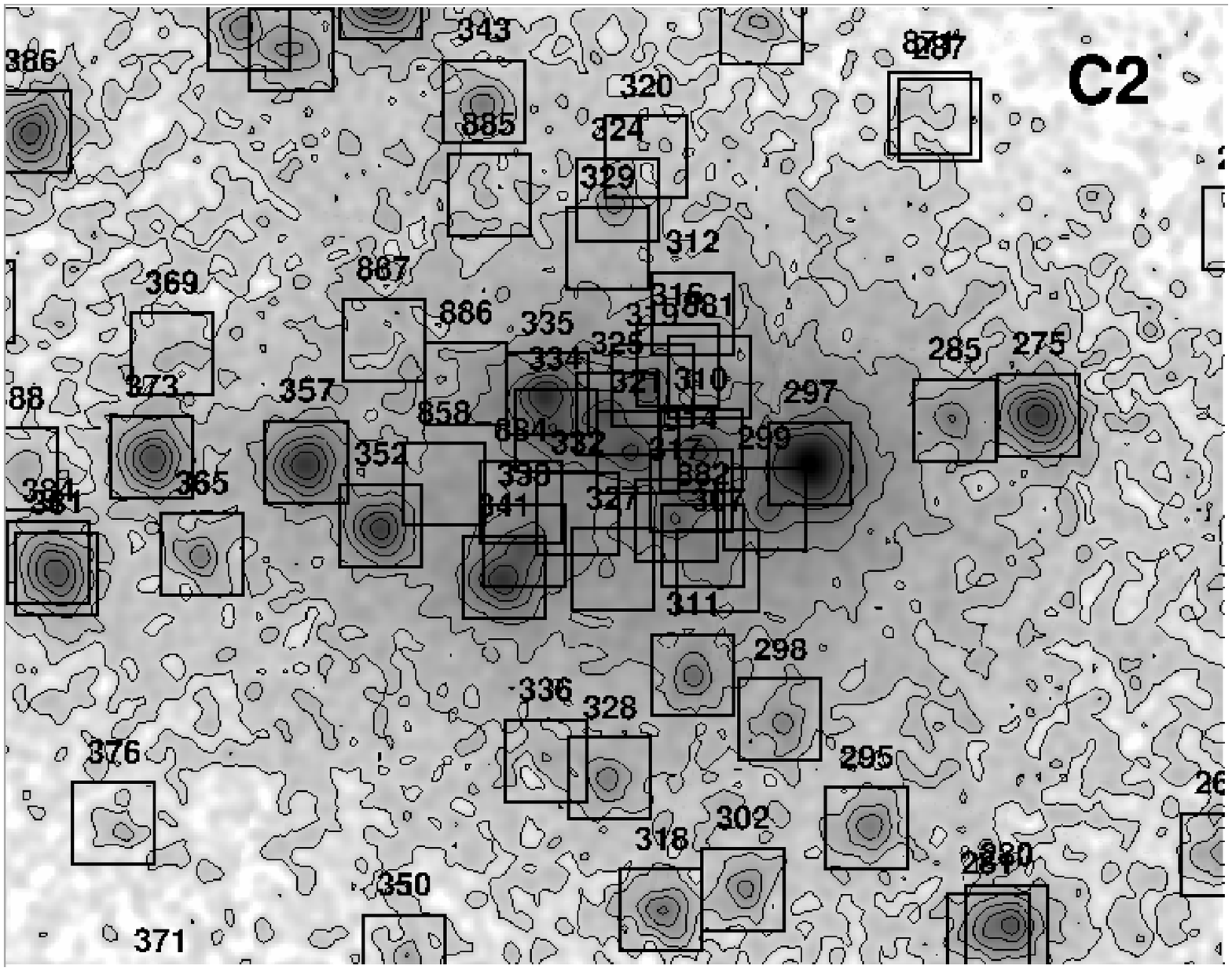}}
  
\resizebox{\hsize}{!}{\includegraphics[clip]{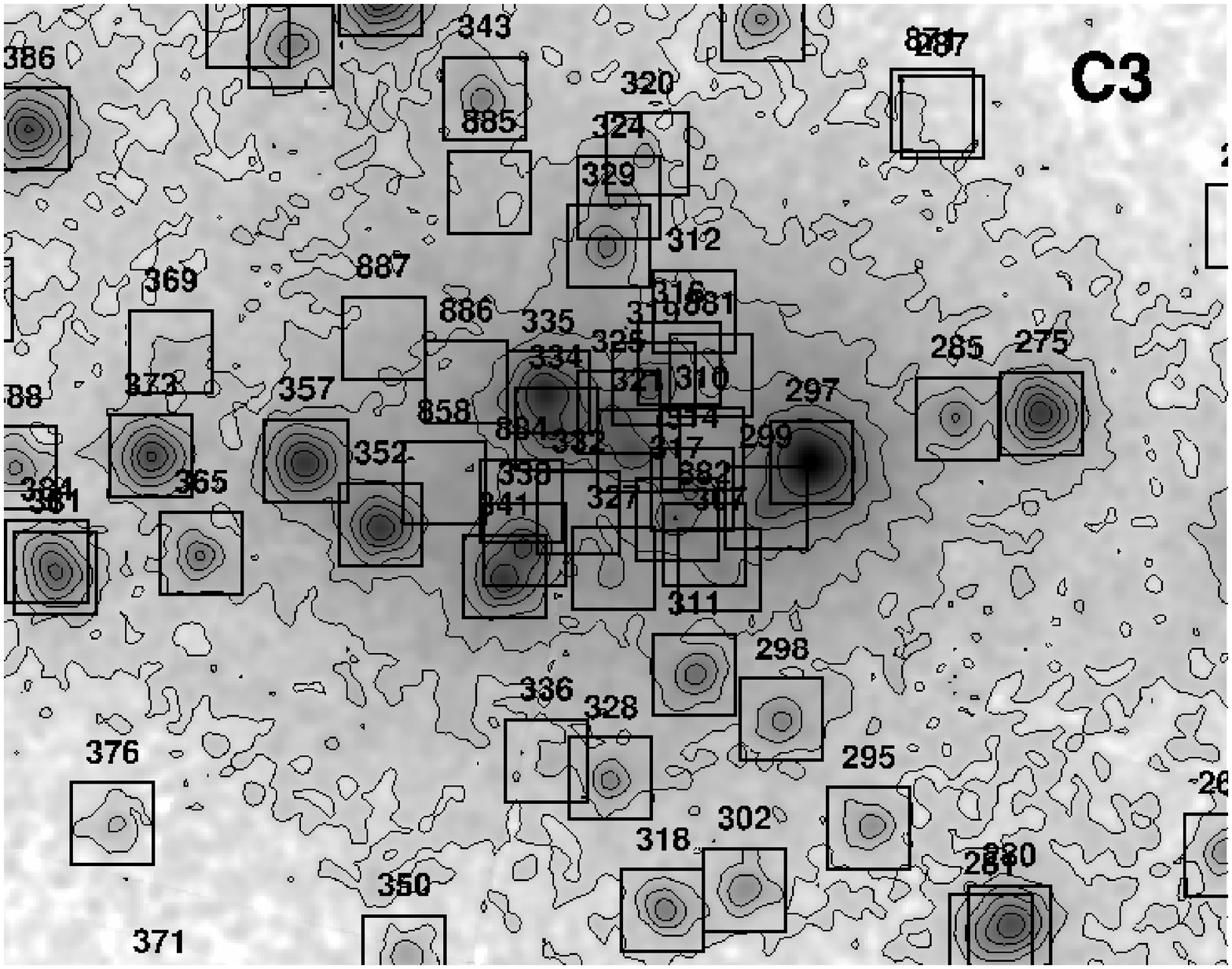}\hskip0.2cm\includegraphics[clip]{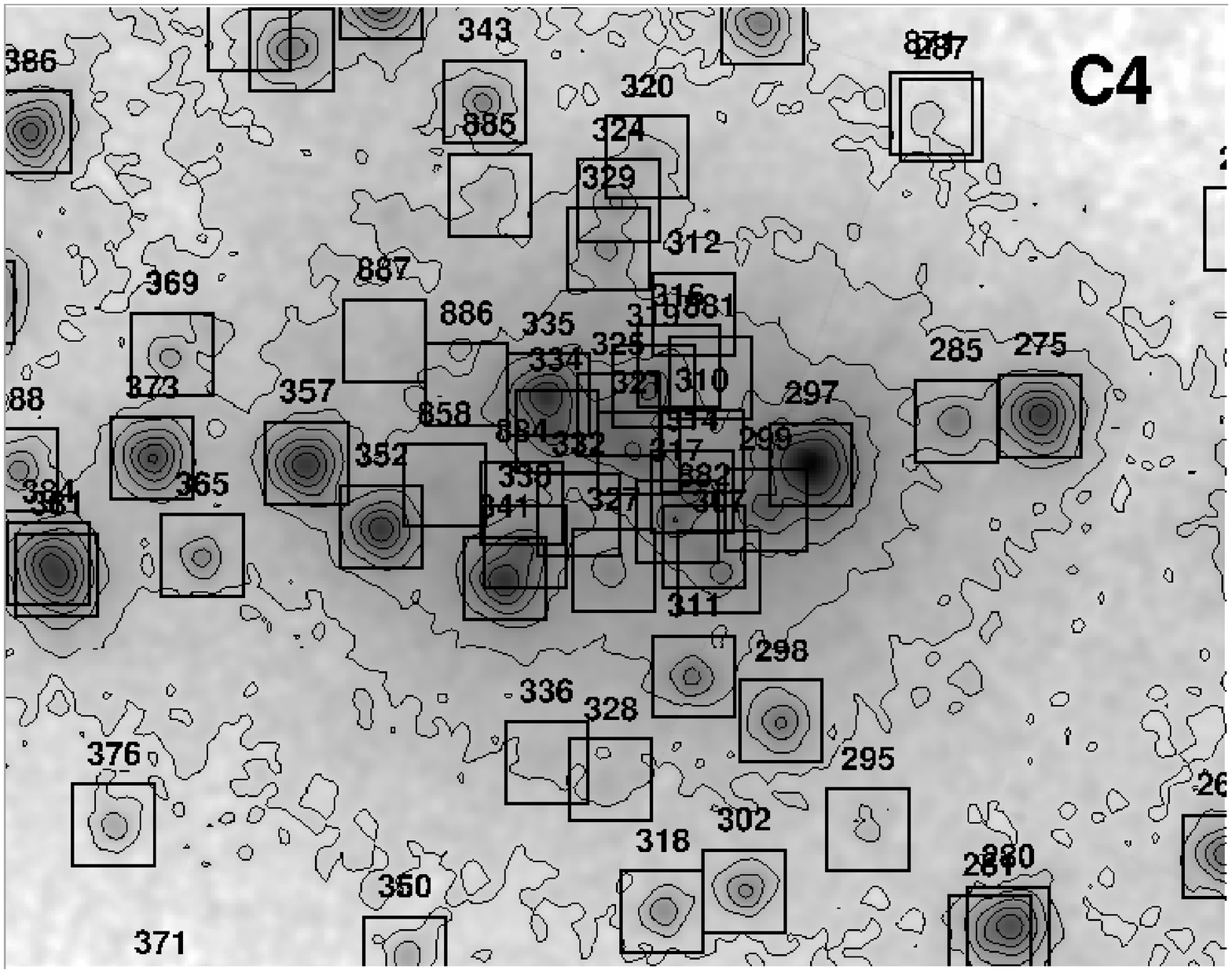}}

\resizebox{\hsize}{!}{\includegraphics[clip]{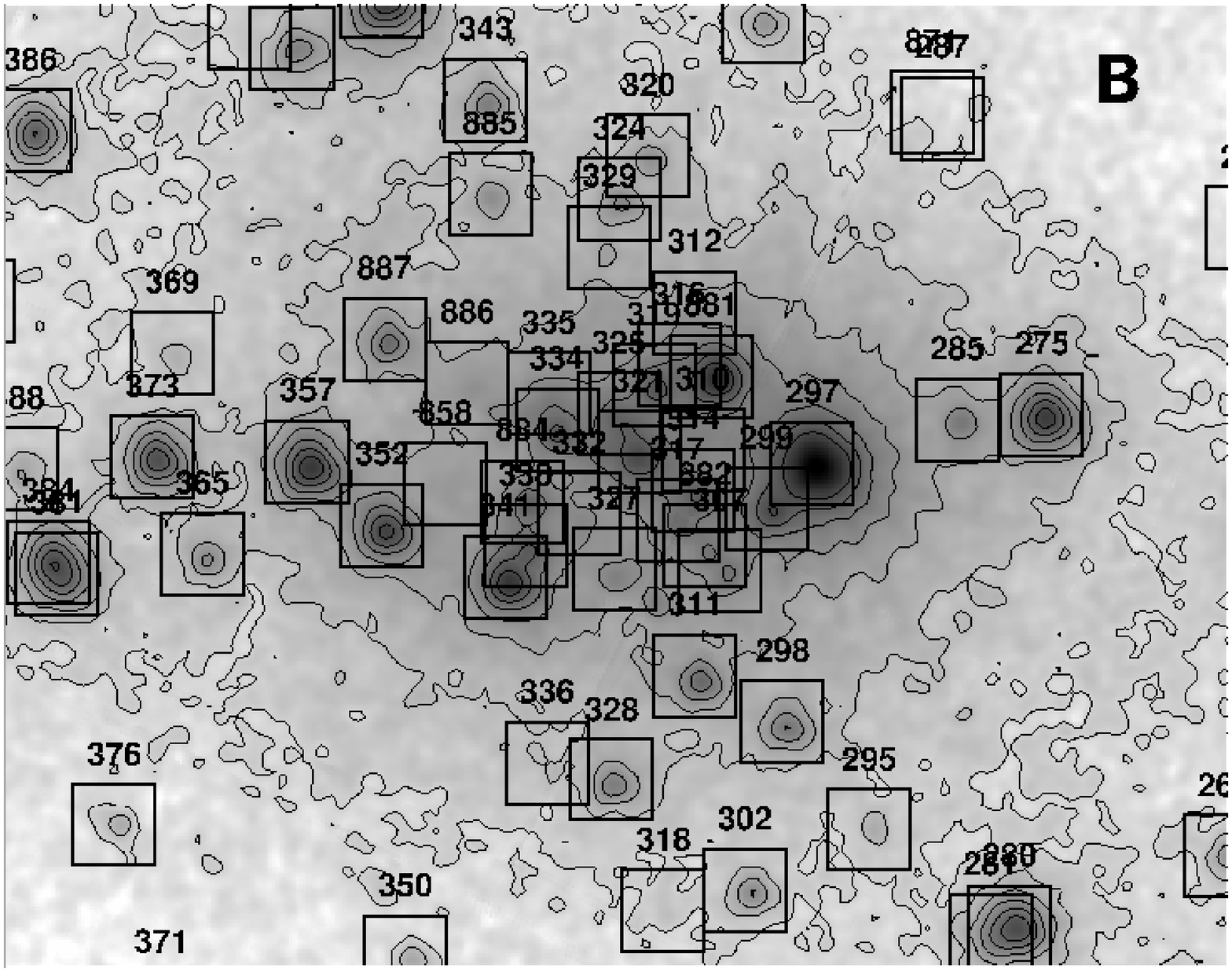}\hskip19.2cm}

     \caption[]{Inner area of \m31\ enlarged from Fig.~\ref{ima_centre}. Contours are at $(4, 8, 16,
     32, 64, 128, 256)\times 10^{-6}$ ct s$^{-1}$ pix$^{-1}$ including a 
     factor of two smoothing. Sources from the combined catalogue are marked as
     30\arcsec$\times$30\arcsec\ squares. The images are ordered as follows:
     Centre 1 ({\bf upper left}),
     Centre 2 ({\bf upper right}),
     Centre 3 ({\bf middle left}),  
     Centre 4 ({\bf middle right}) and
     Centre B ({\bf lower left}).
}
    \label{ima_zoom} 
\end{figure*}

\section{Source catalogue}
\label{srccat}

PFH2005 reported 265 sources in the centre field of \m31. Our catalogue extension contains 39 sources. Four are detected in observation c1, eight in observation c3, thirteen in c4 and twenty one in ``b".

The source parameters are summarised in Table~3 (EPIC combined products and products for EPIC PN, MOS1 and MOS2, separately). 

With the exception of the newly added \xmm\ source name (column 77, see below) Table~3 is structured in the same way as Table~2 from PFH2005. It gives the source number (Col.~1), detection field, from which the source was entered into the catalogue extension (2), source position (3 to 9) with $1\sigma$ uncertainty radius (10), likelihood of existence (11), integrated PN, MOS1 and MOS2 count rate and error (12,13) and flux and error (14,15) in the
(0.2--4.5) keV XID band, and hardness ratios and errors (16--23). Hardness ratios are calculated only for sources for which at least one of the two band count rates has a significance greater than $2\sigma$. Errors are the properly combined statistical errors in each band and can extend beyond the range of allowed values of hardness ratios as defined previously (--1.0 to 1.0). 
The EPIC instruments contributing to the source detection, are indicated in the ``Val" parameter (Col. 24, first character for PN, second MOS1, third MOS2) as ``T", if inside the FOV, or ``F", if outside of FOV. 
There are 8 sources at the periphery of the FOV where only part of the EPIC 
instruments contribute. The positional error (10) does not include intrinsic systematic errors which amount to 0\farcs5 (see PFH2005) and should be quadratically added to the statistical errors. 

Table~3 then gives for EPIC PN, exposure (25), source existence likelihood (26),  count rate and error (27, 28) and flux and error (29, 30) in the
(0.2--4.5) keV XID band, and hardness ratios and errors (31--38). Columns 39 to 52 and 53 to 66 give the same information corresponding to Cols. 25 to 38, but for the EPIC MOS1 and MOS2 instruments. Hardness ratios for the individual instruments were again screened as described above. From the comparison of the 
hardness ratios derived from integrated PN, MOS1 and MOS2 count rates (Cols. 16--23) and the hardness ratios of the individual instruments (Cols. 31--38, 45--52 and 59--66) it is clear that combining the instrument count rate information yielded significantly more hardness ratios above the chosen significance threshold.

Column 67 shows cross correlations with \m31\ X-ray catalogues in the literature. 

Our catalogue extension contains 23 until now unknown X-ray sources in \m31. The discussion of the results of the cross correlation is in Sect.~\ref{diss}.

In the remaining columns of Table~3, we give cross correlation information with sources in other wavelength ranges.

\begin{table*}
\begin{center}
\caption[]{Summary of identifications and classifications.}
\begin{tabular}{llrr}
\hline\noalign{\smallskip}
\hline\noalign{\smallskip}
\multicolumn{1}{c}{Source type$^{\dagger}$} & 
\multicolumn{1}{c}{Selection criteria} &
\multicolumn{1}{c}{Identified} &
\multicolumn{1}{c}{Classified}  \\ 
\noalign{\smallskip}\hline\noalign{\smallskip}
fg~Star & ${\rm log}({{f}_{\rm x} \over {f}_{\rm opt}}) < -1.0$ and $HR2 - EHR2 < 0.3$ and $HR3 - EHR3 < -0.4$ or not defined &  & 1 \\
AGN  &  Radio source and not classification as SNR from $HR2$ or optical/radio &  &  \\
Gal  &  optical id with galaxy  &  &  \\
GCl  &  X-ray extent and/or spectrum & &\\
SSS  &  $HR1 < 0.0$, $HR2 - EHR2 < -0.99$ or $HR2$ not defined, $HR3$, $HR4$ not defined &   &  3 \\
SNR  &  $HR1 > -0.1$ and $HR2 <-0.2$ and not a fg~Star, or id with optical/radio SNR   & 1 & 6 \\
GlC  &  optical id &  & 1 \\
XRB  &  optical id or X-ray variability  & 3 & 4 \\
hard &  $HR2 - EHR2 > -0.2$ or only $HR3$ and/or $HR4$ defined, and no other classification&  & 15 \\
\noalign{\smallskip}
\hline
\noalign{\smallskip}
\end{tabular}
\label{class}
\end{center}
Notes:\\
$^{ {\dagger}~}$: fg~Star: foreground star, AGN: active galactic nucleus, Gal: galaxy, GCl: galaxy cluster, SSS: supersoft source, SNR: supernova remnant, GlC: globular cluster, XRB: X-ray binary 
\end{table*}
To identify the X-ray sources in the \m31\ field we searched for correlations around the X-ray source positions within the 3 $\sigma$ total X-ray error in the SIMBAD and NED archives and within several catalogues. In columns 68 to 73 of Table~3, we give extraction information from the USNO-B1  catalogue (name, number of objects within search area, distance, B2, R2 and I magnitude of  the brightest object).\@ To improve the reliability
of identifications we used the B and R magnitudes to calculate   
\begin{equation}
{\rm log}({{\rm f}_{\rm x} \over {\rm f}_{\rm opt}}) = {\rm log}({f}_{\rm x}) + ({m}_{B2} + {m}_{R2})/(2\times2.5) + 5.37,
\label{fxopt}
\end{equation}
following \citet[][ see Col. 74]{1988ApJ...326..680M}.

The X-ray sources in the catalogue extension are identified or classified based on properties in the X-ray  (HRs, variability) and of correlated objects in other  wavelength regimes (Table~3, Cols. 75, 76).\@ The criteria are summarised in Table~\ref{class}.\@ Identification and classification criteria are discussed in detail in Sect.~6 of PFH2005. As we have no clear hardness ratio criteria to select XRBs, Crab-like supernova remnants (SNRs) or AGNs we introduced a class $<$hard$>$ for those sources. If such a source shows strong variability (i.\,e.~ V $\ge 10$) on the examined time scales it is likely to be an XRB. 
Fifteen sources are classified as $<$hard$>$. Five sources remain unidentified or without classification. 

The last column (77) of Table~3 contains the \xmm\ source name as registered to the IAU Registry. Source names consist of the acronym XMMM31 and the source position as follows: XMMM31~Jhhmmss.s+ddmmss, where the right ascension is given in hours~(hh), minutes~(mm) and seconds~(ss.s) truncated to decimal seconds and the declination is given in degrees~(dd), arc minutes~(mm) and arc seconds~(ss) truncated to arc seconds, for equinox 2000.   

Only two sources from our catalogue extension (869, 863) are found as extended sources (see Table~\ref{extsrcs} and Sect.~\ref{diss}).
\begin{table}
\addtocounter{table}{+1} 
\begin{center}
\caption[]{Extension properties of sources 863 and 869}
\begin{tabular}{rrrr}
\hline\noalign{\smallskip}
\hline\noalign{\smallskip}
\multicolumn{1}{c}{Source} & 
\multicolumn{1}{c}{Extent} &
\multicolumn{1}{c}{Ext. err.$^{\dagger}$} &
\multicolumn{1}{c}{MELH$^{\ddagger}$}  \\ 
\noalign{\smallskip}
& \multicolumn{1}{c}{arcsec$^*$} &
\multicolumn{1}{c}{arcsec$^*$} &  \\ 
\noalign{\smallskip}\hline\noalign{\smallskip}
863 & 6.71 & 2.14 & 4.70 \\
869 & 6.39 & 1.12 & 5.05 \\
\noalign{\smallskip}
\hline
\noalign{\smallskip}
\end{tabular}
\label{extsrcs}
\end{center}
Notes:\\
$^{ {\dagger}~}$: Extent error \\
$^{ {\ddagger}~}$: Maximum extent likelihood\\
$^{ *~}$: $1"$ corresponds to 3.8 pc at the assumed distance of \m31
\end{table}

\section{Variability}
Table~5 contains all information necessary to examine time variability. The sources are taken from the combined catalogue (i.\,e.~PFH2005 and Sect.~\ref{srccat}).\@ Sources are only included in the table, if they are in the FOV for at least two observations. Column 1 gives the source number. Columns 2 and 3 contain the flux and error in the (0.2--4.5) keV XID band. The hardness ratios and errors are given in columns 4 to 11. Column 12 shows cross correlations with \m31\ X-ray catalogues in the literature. The next two columns contain the type of the source (13) and cross correlation information with sources in other wavelength ranges (14).\@ The EPIC instruments contributing to the source detection in the c1 observation, are indicated in the ``c1\_val" parameter (Col. 15, first character for PN, second MOS1, third MOS2) as ``T", if inside the FOV, or ``F", if outside FOV.\@ Then the count rate and error (16,17) and flux and error (18,19) in the (0.2--4.5) keV XID band, and hardness ratios and error (20--27) of the c1 observation are given. Corresponding information is given for observation c2 (cols.~28--40), c3 (41--53), n1 (54--66), c4 (67--79), s1 (80--92) and b (93--105).

Column 106 indicates the number of observations in which the source is covered in the combined EPIC FOV. 
The maxima of the significance of variation and flux ratio (fvar\_max) are given in columns 107 and 108.\@ As described in Sect.~2.3 we only used detections with a significance greater 3 $\sigma$. Otherwise the 3 $\sigma$ upper limit was used. Column 109 indicates the number of observations where we could only gain an upper limit. The maximum flux (fmax) and its error are given in columns 110 and 111. In a few cases we could not derive the maximum flux, because every observation only gives an upper limit. This can have two reasons: The first reason is that PFH2005 merged observations c1 to c4 for source detection. Hence a faint source may not be detectable at the 3$\sigma$ limit in the individual observations. The second reason is, that in cases where the significance of detection was not much above the 3 $\sigma$ limit, it can become smaller than the 3 $\sigma$ limit when the source position is fixed. The source name, according to the IAU naming convention (see Sect.~\ref{srccat}), can be found in column 112.\\     

\begin{figure}
   \resizebox{\hsize}{!}{\includegraphics[clip,angle=-90,bb=52 90 355 460]{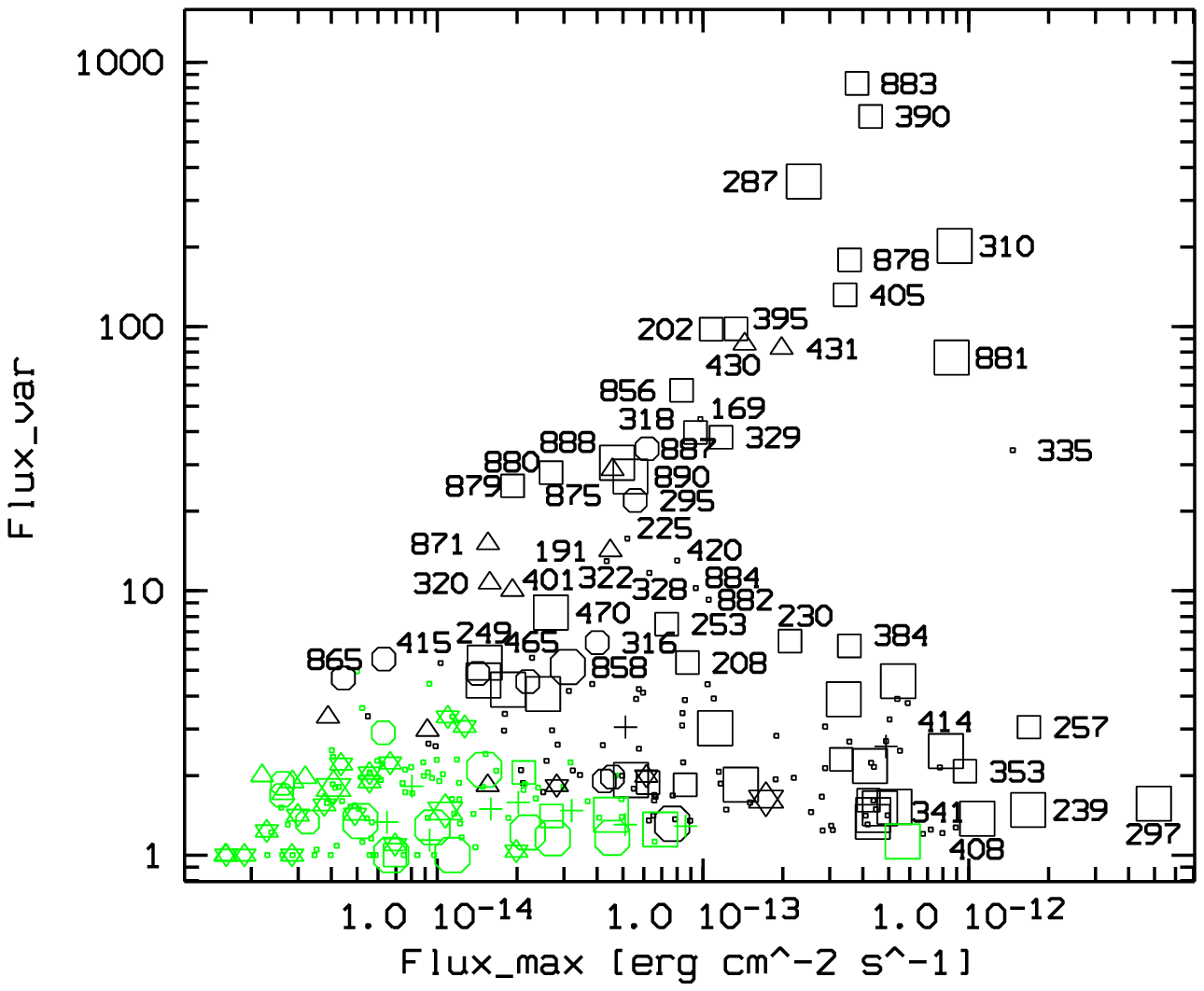}}
     \caption[]{
     Variability factor of \m31\ centre sources of PFH2005 and Sect.~3 in the 0.2--4.5 keV band derived from average fluxes of the \xmm\ EPIC observations from June 2000 to July 2004 plotted versus maximum detected flux (\hbox{erg cm$^{-2}$ s$^{-1}$}). Source classification from PFH2005 is indicated: Foreground stars and candidates are marked as big and small stars, AGN candidates as small crosses, SSS candidates as triangles, SNR and candidates as big and small hexagons, GlCs and XRBs and candidates as big and small squares. Sources with a statistical significance for the variability below 3 are marked in green (grey). Source numbers from PFH2005 and Sect.~3 are indicated for sources with flux variability above 5 or maximum flux above 8\ergcm{-13}.
     }
    \label{var_fmax} 
\end{figure}

In Fig.~\ref{var_fmax} we plotted the variability factor (col.~fvar\_max) of each source as function of its maximum flux (col.~fmax) in the XID band. Identified sources are marked with big symbols, whereas classified sources are indicated by small symbols. Source numbers from PFH2005 and Sect.~\ref{srccat} are indicated for sources with flux variability above 5 or maximum flux above 8\ergcm{-13}.\@ In this region only, variability can help distinguish between foreground stars or SNRs, or to decide if a source classified as hard is an AGN or a XRB.\@ Sources with a statistical significance of the variability below 3 are marked in green (grey).    

\begin{figure*}
   \resizebox{\hsize}{!}{\includegraphics[clip,angle=-90,bb=50 85 360 490]{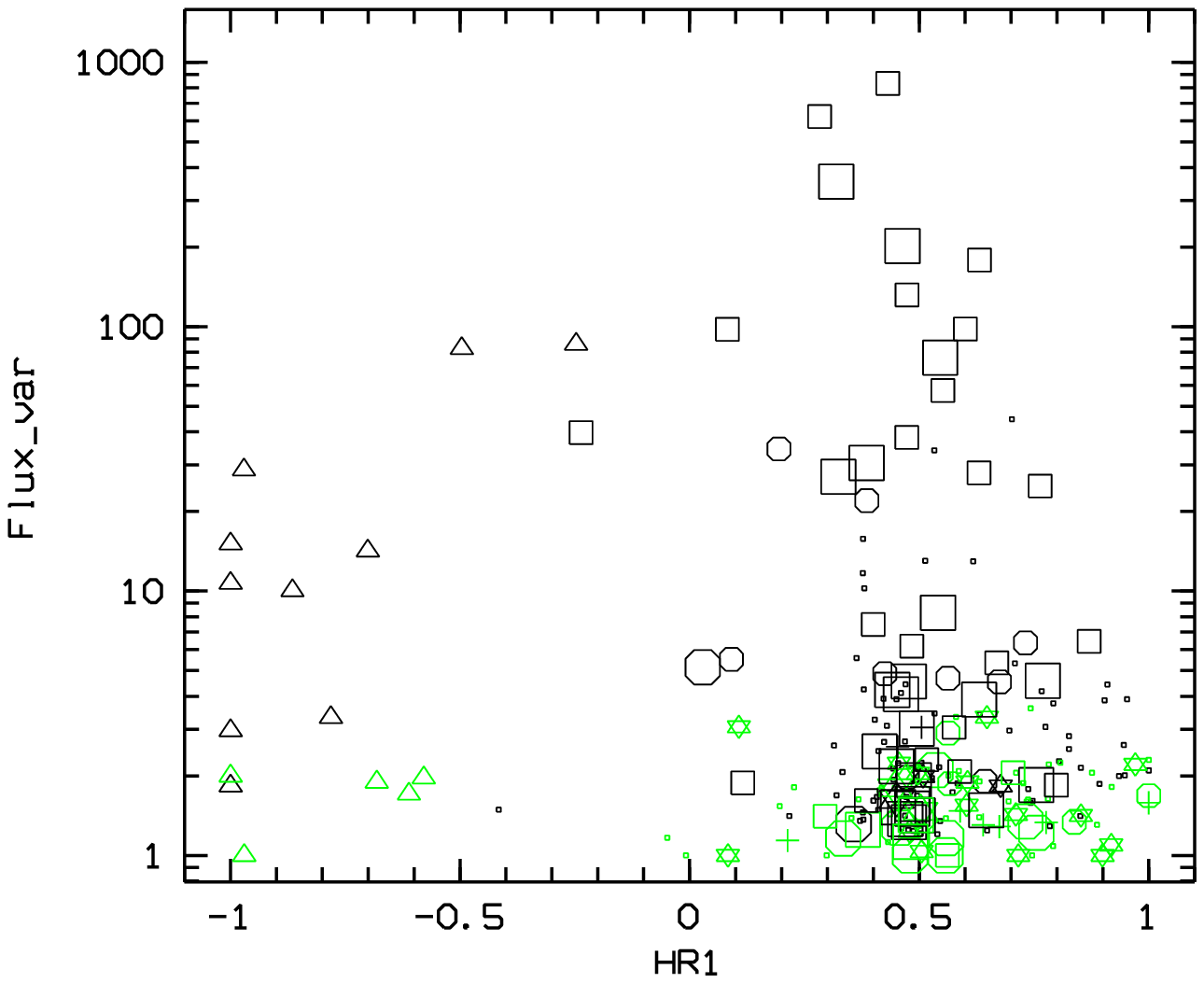}\hskip0.2cm\includegraphics[clip,angle=-90,bb=50 85 360 490]{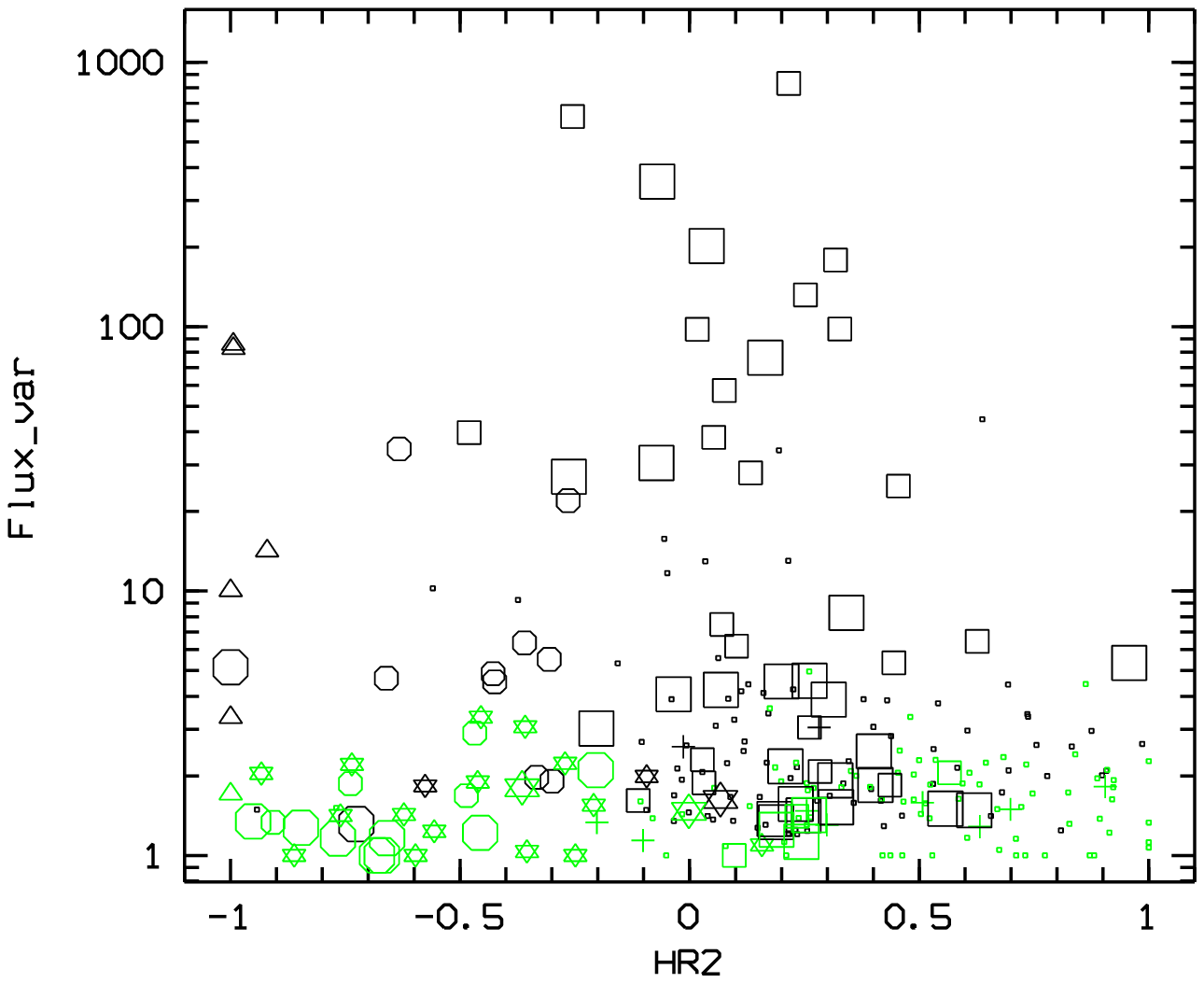}}
     \caption[]{
     Variability factor of \m31\ centre sources of PFH2005 and Sect.~3 in the 0.2--4.5 keV band comparing average fluxes of the \xmm\ EPIC observations from June 2000 to July 2004 plotted versus HR1 in the left panel and HR2 in the right panel, respectively. For source classification see Fig.~\ref{var_fmax}. Sources with a statistical significance of the variability below 3 are marked in green (grey). 
     }
    \label{var_hr} 
\end{figure*}

\begin{table*}
\addtocounter{table}{+1}
\begin{center}
\caption[]{Variable sources with flux variability larger than 5, ordered by variability.}
\begin{tabular}{llrrrcl}
\hline\noalign{\smallskip}
\hline\noalign{\smallskip}
\multicolumn{1}{c}{Source} & 
\multicolumn{1}{c}{Name} &
\multicolumn{1}{c}{fvar} &
\multicolumn{1}{c}{svar} &
\multicolumn{1}{c}{fmax$^{\ddagger}$} &
\multicolumn{1}{c}{type$^{+}$} &
\multicolumn{1}{c}{Comment$^{\dagger}$}  \\ 
\noalign{\smallskip}
& \multicolumn{1}{c}{XMMM31~J} & & & & & \\
\noalign{\smallskip}\hline\noalign{\smallskip}
883 & 004247.8+411113 & 831.10 & 54.75 & 38.02 & $<$GlC$>$ & 1(r), 2(t, 92.2), 12, 17\\
390 & 004305.7+411703 & 624.05 & 79.83 & 42.71 & $<$XRB$>$ & 1(t), 2(t, 954.2), 3(t, 2163), 20, 21(t), 23  \\   
287 & 004234.3+411810 & 353.67 & 43.26 & 23.95 &    XRB    & 1(t), 2(t, 370.5), 15(t, BH-XRN), 21(t), 22(v,t) \\
310 & 004242.1+411608 & 201.71 & 62.44 & 88.47 &    XRB    & 1(t), 2(t, 468.8), 3(t, 285), 15(v,t, BH-XRT), 19(t), 22(v,t) \\   
878 & 004144.7+411111 & 178.79 & 40.20 & 35.61 & $<$XRB$>$ & 1(t,sv), 4(t) \\
405 & 004309.8+411900 & 131.73 & 57.22 & 34.25 & $<$XRB$>$ & 1(sv, $<$AGN$>$), 2(t, 96.3), 3(t, 107), 10, 12(v), 13, 14(v), 20, 22(v,t) \\
202 & 004205.8+411329 &  97.89 & 22.41 & 13.32 & $<$XRB$>$ & 1(r), 2(t, 20.8), 3(t, 93), 12, 15(t), 21(t) \\
395 & 004307.1+411810 &  97.65 & 25.27 & 10.75 & $<$XRB$>$ & 1(t), 2(t, 46.1), 3(t, 155), 20, 21(t), 24  \\
430 & 004318.8+412017 &  85.93 & 40.54 & 14.35 & $<$SSS$>$ & 1(r), 2, 3(t, 96), 10, 13, 14(v), 15(v), 20(v), 22(v) \\
431 & 004319.5+411756 &  82.68 & 40.17 & 19.76 & $<$SSS$>$ & 1(t), 3(t, 694), 15(v,t), 21(t) \\
881 & 004241.8+411635 &  76.27 & 79.68 & 86.17 &    XRB    & 1(t,r,sv), 4(t), 6(t, LMXB), 10, 22(v) \\
856 & 004256.7+411843 &  57.41 & 12.69 &  8.32 & $<$XRB$>$ & 2(t, 79.0), 3(t, 260), 15(t), 19, 21(t), 22(v,t) \\
169 & 004143.4+412118 &  44.67 & 20.22 &  9.77 & $<$XRB$>$ & former type: $<$hard$>$; 1, 14, 24\\
887 & 004252.4+411649 &  39.69 & 20.28 &  9.39 & $<$XRB$>$ & 2(t, 64.6)\\
329 & 004245.1+411723 &  38.07 & 24.24 & 11.71 & $<$XRB$>$ & 1(r,sv), 2(t, 99.5), 3(t, 158), 22(v,t) \\
318 & 004243.3+411319 &  34.47 & 21.30 &  6.17 &           & former type: $<$SNR$>$; 2(t), 20, 22, 24 \\
335 & 004247.1+411629 &  34.02 & 84.69 &146.44 & $<$XRB$>$ & former type: $<$hard$>$; 1(sv), 2, 10, 12, 13, 14, 20, 22(v)\\
888 & 004309.9+412332 &  30.50 & 18.26 &  4.75 &    XRB    & 1(t), 2(t), 4(t), 7(t, LMXB) \\
875 & 004318.7+411804 &  28.78 & 10.70 &  4.57 & $<$SSS$>$ & \\
880 & 004233.9+412331 &  28.02 & 10.39 &  2.68 & $<$XRB$>$ & 2(t, 65.2)\\
890 & 004315.4+412440 &  27.06 & 20.07 &  5.34 &    XRB    & 1(t), 4(t) \\
879 & 004224.5+412401 &  24.95 &  9.75 &  1.92 & $<$XRB$>$ & 2\\
295 & 004236.7+411349 &  21.97 & 11.77 &  5.56 & $<$fgStar$>$& former type: $<$SNR$>$; 2, 13, 14, 20, 22, 24\\
225 & 004210.9+410647 &  15.76 &  8.30 &  5.20 & $<$XRB$>$ & former type: $<$hard$>$; 2, 22(v)\\
871 & 004234.6+411812 &  15.09 &  3.67 &  1.55 & $<$SSS$>$ & 18\\
191 & 004154.3+410724 &  14.18 & 14.03 &  4.48 & $<$SSS$>$ & 1(t) \\
420 & 004316.0+411842 &  13.01 & 17.55 &  7.99 & $<$XRB$>$ & former type: $<$hard$>$; 1, 2, 13, 20, 22(v)\\
322 & 004244.2+412809 &  12.96 & 11.83 &  4.34 & $<$XRB$>$ & former type: $<$hard$>$; 1, 2, 13, 14\\
328 & 004245.0+411407 &  11.69 & 19.22 &  6.29 & $<$XRB$>$ & former type: $<$hard$>$; 1, 2, 12 ,13, 20, 22\\
320 & 004243.8+411756 &  10.70 & 13.89 &  1.58 & $<$SSS$>$ & 3(t, 51), 20 , 24\\
884 & 004247.9+411549 &  10.24 & 25.85 &  9.41 &           & 2, 20, 22, 24\\
401 & 004308.5+411820 &  10.05 & 12.08 &  1.92 & $<$SSS$>$ & 3(t, 38) \\
882 & 004242.0+411533 &   9.26 & 14.38 & 10.50 &           & 2, 20, 22(v)\\
470 & 004336.6+410812 &   8.27 &  8.54 &  2.68 &    GlC    & 5 \\
253 & 004221.6+411418 &   7.48 & 17.56 &  7.31 & $<$GlC$>$ & 1(sv,burst), 2, 8, 20 , 22(v)\\
230 & 004212.1+411757 &   6.45 & 27.71 & 21.25 & $<$GlC$>$ & 1(sv), 2, 5, 12, 15(v),16, 20, 22(v)  \\
316 & 004242.8+411639 &   6.37 &  8.01 &  4.00 &           & former type: $<$SNR$>$; 10, 12(v), 20, 22(v)\\
384 & 004303.3+411527 &   6.19 & 37.24 & 35.54 & $<$XRB$>$ & 1(sv), 2(t, 58.6), 3(t, 33), 5, 10, 12(v), 13, 14, 20, 22(v,t)\\
465 & 004333.4+412140 &   5.58 &  5.13 &  2.28 & $<$hard$>$& 2\\
865 & 004323.4+412208 &   5.52 &  3.57 &  0.63 & $<$SNR$>$ & \\
208 & 004207.0+410017 &   5.35 &  5.10 &  8.75 & $<$GlC$>$ & 5, 13, 14, 16, 21 \\
249 & 004219.6+412153 &   5.35 & 10.47 &  1.51 &    GlC    & 2, 5, 16, 22\\
415 & 004314.5+411649 &   5.33 &  5.21 &  1.03 &           & 2, 22, 24 \\
858 & 004250.4+411556 &   5.14 &  9.69 &  3.10 &    SNR    & 2, 9, 22, 24 \\

\noalign{\smallskip}
\hline
\noalign{\smallskip}
\end{tabular}
\label{varlist}
\end{center}                               
Notes:\\
$^{ {\ddagger}~}$: maximum flux in units of 1\ergcm{-14} or maximum luminosity in units of 7.3\ergs{35}\\
$^{ {+}~}$: type according to Table~\ref{class}, partly changed as mentioned in the comment column\\
$^{ {\dagger}~}$: 1: \citet{2006astro.ph.10809T}, 2: \citet{2007A&A...468...49V}, 3: \citet{2006ApJ...643..356W}, 4: \citet{2006ApJ...645..277T}, 5: \citet{2004ApJ...616..821T}, 6: \citet{2006ApJ...637..479W}, 7: \citet{2005ApJ...632.1086W}, 8: \citet{2005A&A...430L..45P}, 9: \citet{2003ApJ...590L..21K}, 10: \citet{1991ApJ...382...82T}, 11: \citet{1990ApJ...356..119C}, 12: \citet{1993ApJ...410..615P}, 13: \citet{1997A&A...317..328S}, 14: \citet{2001A&A...373...63S}, 15: \citet{2001A&A...378..800O}, 16: \citet{2002ApJ...570..618D}, 17: \citet{2005PASP..117.1236F}, 18: \citet{2007A&A...465..375P}, 19: \citet{2000ApJ...537L..23G}, 20: \citet{2002ApJ...578..114K}, 21: \citet{2004ApJ...609..735W}, 22: \citet{2002ApJ...577..738K}, 23: \citet{2005ApJ...620..723W}, 24: \citet{2004ApJ...610..247D}, 25: \citet{2003A&A...411..553B}; t: transient, v: variable, sv: spectrally variable, r: recurrent, d: dipping, z: Z-source candidate; BH: black hole, XRN: X-ray nova, XRT: X-ray transient, LMXB: low mass X-ray binary, NS: neutron star; numbers indicate the variability given by the corresponding paper 
\normalsize                                                                                                                                                                                                                                                                                                      
\end{table*}                                                                                                                                                                                                                                                                                                     
                                                                                                                                                                                                                                                                                                                 
\begin{table*}                                                                                                                                                                                                                                                                                                   
\begin{center}                                                                                                                                                                                                                                                                                                   
\caption[]{Sources with maximum flux larger than 8\ergcm{-13}, a statistical significance of variability larger than 10 and a flux variability smaller than 5, ordered by flux.}                                                                                                                                                                                                                                                                     
\begin{tabular}{llrrrcl}                                                                                                                                                                                                                                                                                            
\hline\noalign{\smallskip}                                                                                                                                                                                                                                                                                       
\hline\noalign{\smallskip}                                                                                                                                                                                                                                                                                       
\multicolumn{1}{c}{Source} &                                                                                                                                                                                                                                                  
\multicolumn{1}{c}{Name} &                                                                                                                                                                \multicolumn{1}{c}{fvar} & 
\multicolumn{1}{c}{svar} & 
\multicolumn{1}{c}{fmax$^{\ddagger}$} &                                                                                         
\multicolumn{1}{c}{type$^{+}$} &
\multicolumn{1}{c}{Comment$^{\dagger}$}  \\
\noalign{\smallskip} 
& \multicolumn{1}{c}{XMMM31~J} & & & & & \\
\noalign{\smallskip}\hline\noalign{\smallskip}
297 & 004238.5+411603 & 1.56 & 47.20 & 49.71 &    XRB    & 1(sv,z), 2, 10(v), 12(v), 13, 14, 20, 22(v), 25(LMXB) \\
257 & 004223.0+411534 & 3.05 & 51.35 & 16.84 & $<$XRB$>$ & 1(sv), 2, 10(v), 12(v), 13, 14, 20(v), 22(v) \\
239 & 004215.7+410115 & 1.48 & 10.70 & 16.73 &    GlC    & 10, 11(v), 12, 13, 14(v), 16, 20, 21(v) \\   
408 & 004310.6+411451 & 1.37 & 12.81 & 10.77 &    GlC    & 1(sv), 2, 5, 10, 12, 13, 14, 16, 20, 22(v) \\
353 & 004252.5+411854 & 2.07 & 29.97 &  9.68 & $<$XRB$>$ & 1(sv), 2, 10, 12, 13, 14, 20(v, NS-LMXB), 22(v) \\
341 & 004248.5+411522 & 1.27 & 10.99 &  8.94 & $<$hard$>$& 1(sv), 2, 10, 12, 13, 14, 20, 22(v,sv)\\
414 & 004314.3+410722 & 2.47 & 26.89 &  8.21 &    GlC    & 1(d,sv), 2(t, 53.4), 5, 10, 12, 13, 14, 16, 20, 22 \\

\noalign{\smallskip}
\hline
\noalign{\smallskip}
\end{tabular}
\label{varlist2}
\end{center}
Notes:\\
$^{ {\ddagger}~}$: maximum flux in units of 1\ergcm{-13} or maximum luminosity in units of 7.3\ergs{36}\\
$^{ {+}~}$: type according to Table~\ref{class}\\ 
$^{{\dagger}~}$: for comment column see Table~\ref{varlist}
\normalsize
\end{table*}

Figure~\ref{var_fmax} clearly shows that most of the variable sources are XRBs or XRBs in GlC or candidates of these source types. In addition there are a few SSS candidates, and even some SNR candidates showing pronounced temporal variability. These SNR candidates are discussed in Sect.~\ref{diss}, as they should not show time variability. The sources classified or identified as AGNs or foreground stars all show $F_{\mr{var}} < 4$, besides the new foreground star candidate [PFH2005]~295, which is discussed later. 

We found 149 sources with a significance for variability $>3.0$ out of the 300 examined sources. There is a bias towards bright variable sources, because for bright sources it is much easier to detect variability than for faint sources.

Table~\ref{varlist} lists all sources with a variability factor larger than five in descending order. The source number (Col.~1), source name (2), maxima of flux variability (3) and maxima of the significance parameter (4) are given corresponding to Table~5 (Cols.~1, 152, 148 and 147). The next column (5) indicates the type of the source.  If $F_{\mr{var}}\ge10$, sources formerly classified as $<$hard$>$ are now classified as $<$XRB$>$. Time variability can also be helpful to distinguish between foreground star and SNR candidates. In some cases we had to change the source type with respect to PFH2005. This is indicated in the comment column (6). Column 6 also contains references to the individual sources in the literature. In some cases the reference provides information on the temporal behaviour and a more precise type (see brackets). The numbers given in connection with  \citet{2007A&A...468...49V} and \citet{2006ApJ...643..356W} are the \chandra\ derived variability factors obtained in these papers. From the 44 sources listed in Table~\ref{varlist}, six show a flux variability larger than 100. With a flux variability factor $> 830$ source 883 is the most variable source in our sample. Source 335 has the largest significance of variability, with a value of $\approx 85$. Only for ten sources the significance of variability is below 10, for two below 5. Twenty-eight sources are XRBs or XRB candidates and seven are SSS candidates.

Table~\ref{varlist2} lists all ``bright" sources with maximum flux larger than 8\ergcm{-13} and a flux variability smaller than five, giving the same information as in Table~\ref{varlist}. All seven sources listed in Table~\ref{varlist2} have a significance of variability $> 10$.\@ Apart from source 341, they are XRBs (three in globular clusters) or XRB candidates. The most luminous source in our sample is source 297 with a luminosity of $\approx 3.6$\ergs{38}.

Figure~\ref{var_hr} shows the relationship between the variability factor and the hardness ratios HR1 and HR2, respectively. We used the hardness ratios of the observation from which the source entered the catalogue of variable sources. The HR1 plot shows that the sample of highly variable sources includes SSS and XRB candidates, which occupy two distinct regions in this plot \citep[see also ][ for the LMC]{1999A&A...344..521H}. The SSSs marked by triangles, appear on the left hand side, while the XRBs or XRB candidates have much harder spectra, in agreement with their classification. In the HR2 plot the highly variable XRBs and XRB candidates are, apart from the two sources classified as $<$SNR$>$, separated from the bulk of the less variable sources by sources classified as $<$hard$>$. Due to the distinct temporal variability of these sources and the strong absorption in the central region of \m31, it is very unlikely that they are AGNs. So only $<$fg star$>$ or $<$XRB$>$ will be left as possible classification. In accordance with the hardness ratios we suggest sources 169, 225, 322, 328, 335 and 420 as XRB candidates. 
 
Individual sources are discussed in the next section.

\section{Discussion}
\label{diss}
In each of the following subsections, we first discuss the sources described in the catalogue extension (Sect.~\ref{srccat}). In addition we reclassified some sources of PFH2005 based on the results of our time variability study and on recent papers in the literature.

We classified the sources described in the catalogue extension into different types of X-ray emitting objects: foreground stars (fg~Star), galaxies (Gal), AGN, supersoft sources (SSS), supernova remnants (SNR) and X-ray binaries (XRB), using the X-ray properties together with information from catalogues at other wavelengths. The selection criteria for these classes are given in Table~\ref{class}. Additionally we use the time variability to classify sources. In the field of \m31\ mainly XRBs or SSSs can show very strong variability ($F_{\mr{var}} \ge 10$) on time scales of years. In only a few cases we were able to identify an X-ray source with a source already classified from the optical, infrared or radio data. We have no well-defined hardness ratio criteria to differentiate between $<$hard$>$ sources (XRBs, Crab-like SNRs or AGNs).  
Fifteen sources of the catalogue extension are classified as $<$hard$>$ (see Table~\ref{class}).\@ Three of them were found with \chandra\ \citep{2002ApJ...577..738K,2007A&A...468...49V}. Five sources remain unidentified or without classification. Two of the five are already known from \chandra\ observations (see Table~\ref{varlist}).\@ \citet{2002ApJ...577..738K} classified source 884 as SSS.

\subsection{Foreground stars}  
Foreground stars are a class of X-ray sources which is homogeneously distributed
over the field of \m31. The good positioning of \xmm\ and the available catalogues USNO-B1 and 2MASS allow us to effectively select this type of source. We found one foreground star candidate (877) in our source catalogue extension. From the optical colours in the USNO-B1 catalogue we estimate the type to be A3 III or A5 III, using the stellar spectral flux library from \citet{1998PASP..110..863P}.
Another possible foreground star candidate (859) is a USNO-B1 and 2MASS source. From the USNO-B1 magnitudes we derived $f_{\mr{x}}/f_{\mr{opt}}\approx -0.87$ and $f_{\mr{x}}/f_{\mr{opt, R}}\approx -1.27$, where $f_{\mr{opt, R}}$ is the flux in the R-band. The $f_{\mr{x}}/f_{\mr{opt}}$ value is too large, to satisfy our foreground star selection criterion. But for very red objects it can be sufficient that $f_{\mr{x}}/f_{\mr{opt, R}}$ is $<-1$.\@ The source could be a foreground star, in agreement with the values we found for source 295 (see below). But \citet{2007AJ....134..706K} suggested this optical source as possible globular cluster. This classification would also be in agreement with our hardness ratios, $f_{\mr{x}}/f_{\mr{opt}}$ values and USNO-B1 magnitudes \citep[see ][]{2005PASP..117.1236F}. So we cannot decide on a fg~Star or XRB nature and we classify source 859 as $<$hard$>$.

\begin{figure}
 \resizebox{\hsize}{!}{\includegraphics[clip,angle=-90]{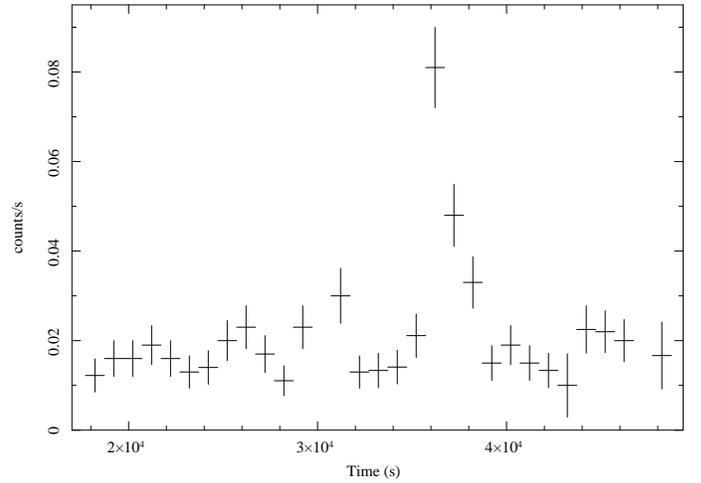}}
 \caption[]{Summed EPIC PN, MOS~1 and MOS~2 $0.2-2.0$ keV light curve of source 295 in the c3 observation binned with 1000s and without background subtraction. The zero time corresponds to 2001-06-29 07:53:36.}
 \label{ima_src295} 
\end{figure}

PFH2005 classified source 295 as a SNR. This classification has to be rejected due to the distinct time variability we found. We created light curves in the $0.2-2.0$ keV range for the different observations. In some, especially in c3 (see Fig.~\ref{ima_src295}) and  in c4, the source shows strong flares. The observation c2 consists of the decaying wing of a strong flare, while the source remains rather quiet in ``b".  
In addition we carefully checked the 2MASS and Local Group (LG) survey R-band images \citep{2006AJ....131.2478M} and found in both images a faint point-like source, at the X-ray position. Eq.~(\ref{fxopt}) gives $f_{\mr{x}}/f_{\mr{opt}}\approx -0.66$ and $f_{\mr{x}}/f_{\mr{opt, R}}\approx -1.28$  using brightnesses from the LG survey photometric catalogue. The $f_{\mr{x}}/f_{\mr{opt}}$ values derived from the catalogue by \citet[][ $f_{\mr{x}}/f_{\mr{opt}}\approx -0.60$ and $f_{\mr{x}}/f_{\mr{opt, R}}\approx -1.44$]{1994A&A...286..725H} are in good agreement with the values derived form the LG Survey and are reasonable for a red star. Considering all those points, we now classify this source as a foreground star.

\subsection{Supersoft sources}
Spectra of SSSs with low energy resolution can be modelled by black body spectra with temperatures below 50 eV. They radiate close to the Eddington luminosity of a 1 $M_{\sun}$ object and are believed to be white dwarf systems steadily burning hydrogen at the surface. They were identified as a class of X-ray sources by ROSAT and are often observed as transient X-ray sources  \citep[see][ and references therein]{2000NewA....5..137G}. 

Our catalogue extension contains three SSSs. Two of them (871, 886) correlate with optical novae and have been investigated in more detail in PFF2005 and PHS2007. 

The third one (875) lies near source [PFH2005]~431 (distance $\approx$ 12"). As source 431 is brightest in observation c1 and source 875 is detected in observation c4, we can exclude that they are the same source. From the time variability and the positional errors it would be possible that source 875 corresponds to the nova M31N1923-12b (= [H29]~N28; distance $\approx$ 7"), which was reported in the optical wavelength regime by \citet[][ see also Nova list of PHS2007]{1929ApJ....69..103H}. However super soft X-ray emission from novae up to now has only been observed up to ten years after the optical outburst (see e.\,g.~PHS2007).
So if source 875 really coincides with M31N1923-12b, the X-ray emission we found would have to be connected with an unreported optical outburst, which occurred during the last ten years, making the source a recurrent nova. Another possibility is, that source 875 corresponds to yet another nova, not reported in the optical. But we cannot exclude that source 875 is not a nova at all.

\subsection{Supernova remnants}
SNRs can be separated into sources where thermal components dominate the X-ray spectrum below 2 keV, and so-called ``plerions" with power law spectra. The former are located in areas of the X-ray colour/colour diagrams which only overlap with foreground stars. If we assume that we have identified all foreground star candidates from the optical correlation and inspection of the optical images, the remaining sources can be classified as SNR candidates using the criteria given in Table~\ref{class}. 

We thus identified six SNR candidates in our catalogue extension. One of them (885) had been previously observed with \chandra\ \citep{2002ApJ...577..738K,2002ApJ...578..114K}, but had not been classified.  
A second source (858) coincides with a source reported as a ring-like extended object from \chandra\ observations, which was also detected in the optical and radio wavelength regimes and identified as SNR \citep{2003ApJ...590L..21K}.

Two sources from our catalogue extension, which are classified as SNRs are listed in Table~\ref{varlist}. Source 858 lies next to source 875, which was first detected in observation b. Therefore the flux of source 858 is underestimated in ``b" and the source appears variable. There is thus no need to change the type of this source. For source 865 we can only gain upper limits for the flux, apart from observation c3 ($L_{\mr{X}}\approx4.6$\ergs{35}), which leads to a significance of variability of only 3.57, not much above the 3 $\sigma$ limit. So the source can still be classified as a SNR candidate, despite the alleged time variability.

\begin{figure}
 \resizebox{\hsize}{!}{\includegraphics[clip,angle=-90]{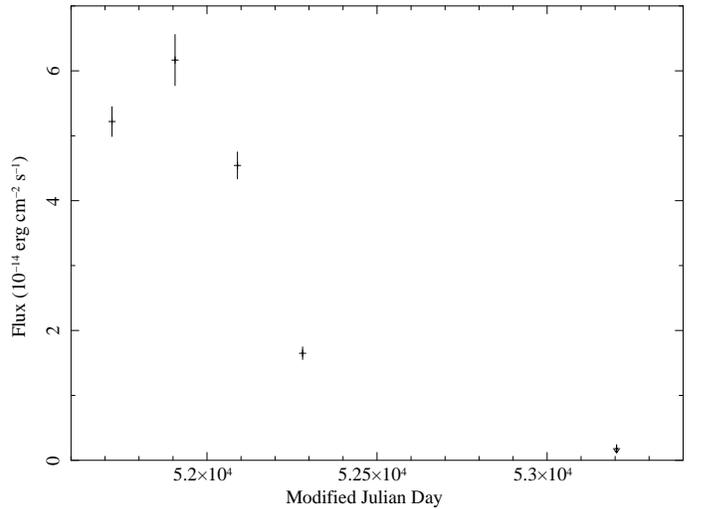}}
 \caption[]{EPIC long-term light curve of source 318. We used XID fluxes. The arrow marks a $3\sigma$ upper limit.}
 \label{ima_src318} 
\end{figure}

We now discuss the SNR candidates of PFH2005, which show time variability: 

Source 318 shows significant variability. Therefore we have to reject the classification of PFH2005 as $<$SNR$>$. Fig.~\ref{ima_src318} shows, that in observation ``b" the source is about a factor of 10 to 35 less luminous than in the other observations. We checked carefully whether the source lies at the rim of a CCD or on a CCD gap. Neither is the case. In the following we discuss possible source classifications: the hardness ratios are in agreement with our foreground star criterion, however, the duration of the outburst of about two years seems much too long for a stellar flare (Fig.~\ref{ima_src318}). Since we also did not find an optical counterpart in the images of the LG survey \citep{2006AJ....131.2478M}, we can exclude a foreground star identification. The behaviour on long-term time scales suggests an X-ray nova as a possible source classification \citep{1999A&A...344..521H, 1996ARA&A..34..607T, 1997ApJ...491..312C}. We used the data of observation c2, in which the source is most luminous, to produce an EPIC PN spectrum. A disk blackbody model fitted to the spectrum gives a temperature at the inner edge of the accretion disk of $\approx 190$ eV, which seems too small for an X-ray Nova or LMXB. We also fitted a blackbody spectrum. The temperature of $\approx 160$ eV is too high for a SSS, but would be in agreement with a QSS \citep{2006ApJ...643..844O,2006ARA&A..44..323F}. A power law fit gives a photon index of $\approx 4.7$. Photon indices of XRBs and AGNs are much smaller than that value. So the nature of this source remains unclear.

For source 316 the variability factor we found is based on observation b (without b: $F_{\mr{var}}=3.07$ and $S_{\mr{var}}=4.75$).\@ As the source lies next to the bright transient source 881, which was first detected in that observation, the flux of source 316 may be underestimated and the source could appear as a variable. However due to the variability reported in the literature (see Table~\ref{varlist}), the SNR classification has to be rejected.

\subsection{Globular cluster sources and X-ray binaries}
A significant part of the luminous X-ray sources in the Galaxy and \m31\ are found in globular clusters. We correlated our catalogue extension with that of \citet{2004A&A...416..917G}. 

All $<$hard$>$ sources of our source catalogue extension, which have a variability factor larger than ten are classified as XRBs. References for these sources can be found in Table~\ref{varlist}.\@ TPC06 report on four bright X-ray transients, which they detected in the observations of July 2004 and suggest them as XRB candidates. We also found these sources and classified source 878 and identified sources 881, 888, 890 as XRBs. One of the identified XRBs (890) shows a very soft spectrum. \citet{2005ApJ...632.1086W} observed source 888 with \chandra\ and \textsl{HST}. From the location and X-ray spectrum they suggest it is a LMXB. They propose as optical counterpart a star within the X-ray error box, which shows a change in optical brightness ($\Delta B$) of $\approx 1$ mag. 
Source 881 was first detected in January 1979 by TF91 with the \ein\ observatory. WGM06 rediscovered it in \chandra\ observations from 2004. Their coordinated \textsl{HST} ACS imaging does not reveal any variable optical counterpart. From the X-ray spectrum and the lack of a bright star WGM06 suggest this source as a LMXB with a black hole.

In PFH2005, sources 169, 225, 322, 328, 335 and 420 were classified as $<$hard$>$. We found that they all have a time variability factor larger than ten and therefore re-classified them as XRB candidates. 

Sources 257 and 384 were proposed as stellar mass black hole candidates by \citet{2003A&A...405..505B} and \citet{2004A&A...423..147B}, respectively. Recently, it was shown that the aperiodic variability of these sources has an artificial origin \citep{2007A&A...469..875B}. So there is no longer clear evidence for a black hole nature of these objects \citep{2007A&A...469..873B}. We now classify sources 257 and 384 as XRB candidates, based on their time variability (see Tables~\ref{varlist} and~\ref{varlist2}).

Source 883 is a transient, only detected in July 2004 (obs. b) in our study. It stands out in Fig.~\ref{var_fmax} and Table~\ref{varlist} as the source with the highest variability ($F_{\mr{var}} \approx 830$). The EPIC pn data of source 883 during observation b can be well fitted with an absorbed power law model ($N_{\mr{H}}=1.1\pm 0.2$ \hcm{21}, photon power law index = $1.61 \pm 0.08$ , unabsorbed $0.5 - 8.0$ keV luminosity = 3.7\ergs{37}). The source correlates with the GlC candidate Bo 128 \citep[e.\,g.~][]{2004A&A...416..917G}.\@ Based on its variability, luminosity and absorbed power law spectrum \citet{2006astro.ph.10809T} classify the source as a neutron star XRB candidate (\#\,77 in their list of bright X-ray sources detected in the central part of \m31).\@  During the \chandra\ monitoring of the centre area of \m31\ the transient was detected at a similar luminosity 2 months earlier in May 2004 \citep[source 136 in ][]{2007A&A...468...49V}, most likely during the same outburst.  No additional \chandra\ detections of the source have been reported. No source was detected at the position of this bright transient with the Einstein Observatory 1979/80 (e.g. TF91), during the ROSAT PSPC surveys \citep[Jul 2001, Jul/Aug 2002, Dec 2002/Jan 2003, Jul 2003; see][]{1997A&A...317..328S,2001A&A...373...63S} and during ROSAT HRI observations in Jul 2004 and Jan 2006 (see source catalogues of the pointed HRI observations 1RXH). However, two additional outbursts of the transient were detected with the ROSAT HRI in July 1990 (source 51 in PFJ93) and in Jul/Aug 1995 (see 1RXH). The luminosity derived for these outbursts is remarkably similar to the luminosity of the outburst in 2004 if we assume that the X-ray spectrum of this recurrent transient can always be described by the same model as during the 2004 outburst (see Table~\ref{tab_outb}). 

\begin{table}
\begin{center}
\caption[]{Outbursts of source 883 = [PFJ93]~51 = [VG2007]~136 = [TPC2006]~77}
\begin{tabular}{cccc}
\hline\noalign{\smallskip}
\hline\noalign{\smallskip}
\multicolumn{1}{c}{Satellite} & 
\multicolumn{1}{c}{Time of observation} &
\multicolumn{1}{c}{$L_x$ $^{+}$} &
\multicolumn{1}{c}{Reference$^{\dagger}$}  \\ 
\noalign{\smallskip}\hline\noalign{\smallskip}
ROSAT HRI & Jul 1990 & 4.7 & 1 \\
ROSAT HRI & Jul/Aug 1995 & 4.6 & 2 \\
\chandra\ ACIS-I & May 2004 & 3.3 & 3 \\
\xmm\ EPIC & Jul 2004 & 3.7 & 4, this work \\
\noalign{\smallskip}
\hline
\noalign{\smallskip}
\end{tabular}
\label{tab_outb}
\end{center}
Notes:\\
$^{ {+}~}$: $0.5-8.0$ keV unabsorbed luminosity in units of \oergs{37} for a distance of 780 kpc, assuming $N_{\mr{H}}=1.1$ \hcm{21} and a photon index of 1.6\\
$^{{\dagger}~}$: 1: \citet{1993ApJ...410..615P}, 2: 1RXH catalogue, 3: \citet{2007A&A...468...49V}, 4: \citet{2006astro.ph.10809T}\\
\normalsize
\end{table}
 
\section{Conclusion}
In this paper we present an updated source list of the central area of the bright Local Group spiral galaxy \m31, using the observations from June 2000 to July 2004 available from the \xmm\ archive. We extended the source catalogue by PFH2005, based on the merged images of the observations from 2000 to 2002 by searching sources in the observations of 2004 and reexamining the observations used in PFH2005 individually. To classify or identify more of the sources, we examined their long term time variability.
 
We obtained 39 sources in addition to the 265 reported by PFH2005 in the field. The identification and classification of these sources is based on properties in the X-ray wavelength regime: hardness ratios and temporal variability. In addition, information from cross correlations with \m31\ catalogues in the radio, infra-red, optical and X-ray wavelength regimes are used.

We detected three SSS candidates, one SNR and six SNR candidates, one GlC candidate, three XRBs and four XRB candidates. Additionally we identified one foreground star candidate and classified fifteen sources as $<$hard$>$, which may either be XRBs or Crab-like SNRs in \m31\ or background AGNs. The remaining five sources remain unidentified and without classification. Two sources were found to be extended. One of them was classified as $<$hard$>$. The other stays without classification.

To examine the time variability we calculated the flux or at least an upper limit at the source positions in each observation. We determined the variability factor and significance parameter for each source, comparing the XID flux ratios of the different observations with each other. The time variability helped us to decide if a source classified as $<$hard$>$ in PFH2005 can be an XRB candidate. In addition we could use time variability to distinguish between foreground star and SNR candidates.
      
Six sources of PFH2005, which were classified as $<$hard$>$, show distinct time variability. Based on that variability, their hardness ratios and the strong absorption in the centre of \m31\ we suggest these sources as XRB candidates. The SNR classification from source 295 was changed to foreground star due to the distinct time variability we found and its identification with a faint stellar object. Other SNR classifications (sources 316, 318) were rejected due to time variability of the sources.

To verify our suggested classifications further investigations, including at other wavelengths will be necessary.

\begin{acknowledgements}
This publication makes use of the USNOFS Image and Catalogue Archive
operated by the United States Naval Observatory, Flagstaff Station
(http://www.nofs.navy.mil/data/fchpix/), 
of data products from the Two Micron All Sky Survey, 
which is a joint project of the University of Massachusetts and the Infrared 
Processing and Analysis Center/California Institute of Technology, funded by 
the National Aeronautics and Space Administration and the National Science 
Foundation, of the SIMBAD database,
operated at CDS, Strasbourg, France, 
and of the NASA/IPAC Extragalactic Database (NED) 
which is operated by the Jet Propulsion Laboratory, California 
Institute of Technology, under contract
with the National Aeronautics and Space Administration.
The XMM-Newton project is supported by the
Bundesministerium f\"ur Wirtschaft und Technologie/Deutsches Zentrum
f\"ur Luft- und Raumfahrt (BMWI/DLR, FKZ 50 OX 0001) and the Max-Planck
Society. HS acknowledges support by the
Bundesministerium f\"ur Wirtschaft und Technologie/Deutsches Zentrum
f\"ur Luft- und Raumfahrt (BMWI/DLR, FKZ 50 OR 0405).
\end{acknowledgements}

\bibliographystyle{aa}
\bibliography{./paper,/home/hstiele/data1/papers/my1990,/home/hstiele/data1/papers/my2000,/home/hstiele/data1/papers/my2001,/home/hstiele/data1/papers/catalog,/home/hstiele/data1/papers/my2007}

\end{document}